\documentclass[10pt]{article}	
\usepackage{float}		
\usepackage{times}		
\usepackage{graphicx}	
\usepackage{url}
\usepackage{epstopdf}
\usepackage{amsmath}
\usepackage{amssymb}
\usepackage{multicol,multirow}
\usepackage[font={small}]{caption}
\usepackage{subcaption}
\usepackage{abstract}
\usepackage{paralist} 
\usepackage{xcolor}
\usepackage[document]{ragged2e}
\usepackage{txfonts}
\usepackage{microtype}
\usepackage{enumitem}  
\usepackage{comment} 
\usepackage[T1]{fontenc}
\usepackage{wrapfig}
\usepackage{array}
\usepackage{booktabs}
\usepackage{placeins}
\usepackage{caption}
\usepackage[round]{natbib}
\usepackage{bm}
\usepackage{listings}
\usepackage{hyperref}

\newcommand{\pv}{\mbox{IRAM 30\,m}}
\newcommand{\kt}{\mbox{Kitt Peak}}

\newcommand{\pasp}{PASP}

\newcommand{\aj}{AJ}
\newcommand{\procspie}{Proceedings of the SPIE}
\hypersetup{
    colorlinks,
    linkcolor={blue!70!black},
    citecolor={blue!70!black},
    urlcolor={blue!70!black}
}
 
\newcommand{\degr}{\hbox{\ensuremath{^{\circ}}}}

\DeclareGraphicsExtensions{.pdf,.png,.jpeg}

\def\memonr{2026-L1-01}

\title{\vspace{-1.25cm} 
\textcolor{gray}{\textit{L1 Processing and Validation}}\\
\vspace{0.40cm}
\textbf{Metadata for the Flux Density Calibration of the April 2021 and March 2022 
Event Horizon Telescope Science Data}
} 

\author{ 
C. Romero-Ca\~nizales$^{1,2}$,
J. Y. Koay$^{3}$,
L. Blackburn$^{4,5}$,
L. D. Matthews$^{6}$,
V. Fish$^{6}$,
R. P. J. Tilanus$^{7}$,\\
G. B. Crew$^{6}$,
K. Asada$^{2}$,
M. Bremer$^{8}$,
C.-S. Chang$^{9,10}$,
P. Friberg$^{11}$,
C. Goddi$^{12,13,14}$,\\
G. Keating$^{4}$,
R. P. Keenan$^{15}$,
J. Kim$^{16}$,
T. P. Krichbaum$^{17}$,
D. P. Marrone$^{7}$,
S. Matsushita$^{2}$,\\
I. Mizuno$^{11}$,
J. P. P\'erez-Beaupuits$^{17,18,19}$,
D. W. Pesce$^{4,5}$,
V. Pi\'etu$^{8}$,
V. Ramakrishnan$^{20}$,\\
A. L. Roy$^{17}$,
I. Ru\'{\i}z$^{21}$,
S. S\'anchez$^{21}$,
P. Torne$^{21}$,
M. Jan\ss{}en$^{1}$,\\
D. G. Nair$^{22,23,17}$,
I. Natarajan$^{4,5}$ \&
S. D. von Fellenberg$^{24,17}$
}
\def\affiliations{
\hspace{0.5cm}$^{1}$\textit{Department of Astrophysics \/ IMAPP, Radboud University, P.O. Box 9010, 6500 GL Nijmegen, The Netherlands}\\
\hspace{0.5cm}$^{2}$\textit{Academia Sinica Institute of Astronomy and Astrophysics, No.1, Sec. 4, Roosevelt Rd, Taipei 10617, Taiwan, R.O.C.} \\
\hspace{0.5cm}$^{3}$\textit{Niigata University , 8050 Ikarashi-nino-cho, Nishi-ku, Niigata 950-2181, Japan} \\
\hspace{0.5cm}$^{4}$\textit{Center for Astrophysics $|$ Harvard \& Smithsonian, 60 Garden St., Cambridge, MA 02138, USA} \\
\hspace{0.5cm}$^{5}$\textit{Black Hole Initiative at Harvard University, 20 Garden Street, Cambridge, MA 02138, USA} \\
\hspace{0.5cm}$^{6}$\textit{Massachusetts Institute of Technology Haystack Observatory, 99 Millstone Road, Westford, MA 01886, USA} \\
\hspace{0.5cm}$^{7}$\textit{Steward Observatory and Department of Astronomy, University of Arizona, 933 N. Cherry Avenue, Tucson, AZ 85721, USA} \\
\hspace{0.5cm}$^{8}$\textit{Institut de Radio Astronomie Millim\'{e}trique, 300 rue de la Piscine, 38400, Saint-Martin-d'H\`{e}res, France} \\
\hspace{0.5cm}$^{9}$\textit{Department of Astronomy, University of Geneva, Chemin Pegasi 51, 1290 Versoix, Switzerland} \\
\hspace{0.5cm}$^{10}$\textit{Institute of Physics, Laboratory of Astrophysics, \'Ecole Polytechnique F\'ed\'erale de Lausanne (EPFL), Observatoire de Sauverny, \\
\hspace{0.65cm} CH-1290 Versoix, Switzerland} \\
\hspace{0.5cm}$^{11}$\textit{East Asian Observatory, 660 N. A’ohoku Place, Hilo, HI 96720, USA} \\
\hspace{0.5cm}$^{12}$\textit{Instituto de Astronomia, Geof\'isica e Ci\^encias Atmosf\'ericas, Universidade de S\~ao Paulo, R. do Matão, 1226, S\~ao Paulo, \\ \hspace{0.65cm} SP 05508-090, Brazil} \\
\hspace{0.5cm}$^{13}$\textit{Dipartimento di Fisica, Universit\'a degli Studi di Cagliari, SP Monserrato-Sestu km 0.7, I-09042 Monserrato (CA), Italy} \\
\hspace{0.5cm}$^{14}$\textit{INAF - Osservatorio Astronomico di Cagliari, via della Scienza 5, I-09047 Selargius (CA), Italy} \\
\hspace{0.5cm}$^{15}$\textit{Max Planck Institut f\"ur Astronomie (MPIA), K\"onigstuhl 17, 69117 Heidelberg, Germany} \\
\hspace{0.5cm}$^{16}$\textit{Department of Physics, Korea Advanced Institute of Science and Technology (KAIST), 291 Daehak-ro, Yuseong-gu, Daejeon \\ \hspace{0.65cm} 34141, Republic of Korea} \\
\hspace{0.5cm}$^{17}$\textit{Max Planck Institut f\"ur Radioastronomie (MPIfR), Auf dem H\"ugel 69, 53121 Bonn, Germany} \\
\hspace{0.5cm}$^{18}$\textit{European Southern Observatory, Alonso de C\'ordova 3107, Vitacura, Casilla 19001, Santiago de Chile, Chile} \\
\hspace{0.5cm}$^{19}$\textit{Centro de Astro-Ingenier\'{\i}a, Pontificia Universidad Cat\'olica de Chile, Casilla 306, Santiago, Chile} \\
\hspace{0.5cm}$^{20}$\textit{Signal Processing Research Centre, Tampere University, 33720 Tampere, Finland} \\
\hspace{0.5cm}$^{21}$\textit{Institut de Radioastronomie Millim\'{e}trique (IRAM), Avenida Divina Pastora 7, Local 20, E-18012, Granada, Spain}\\
\hspace{0.5cm}$^{22}$\textit{Institute of Astronomy, University of Cambridge, Madingley Road, Cambridge CB3 0HA, United Kingdom} \\
\hspace{0.5cm}$^{23}$\textit{Astronomy Department, Universidad de Concepci\'on, Casilla 160-C, Concepci\'on, Chile} \\
\hspace{0.5cm}$^{24}$\textit{Canadian Institute for Theoretical Astrophysics, University of Toronto, 60 St. George Street, Toronto, ON M5S 3H8, Canada}
}

\date{\today -- Version 1.0}
\def\memohistory{
\item {Oct 16, 2024 -- Document created}
}
\newif\ifshowhistory  
\begin{document} 

\begin{figure}[!t]
\begin{minipage}[t]{0.49\textwidth}
    \vspace{-0.13\linewidth}{\includegraphics[width=0.6\linewidth]{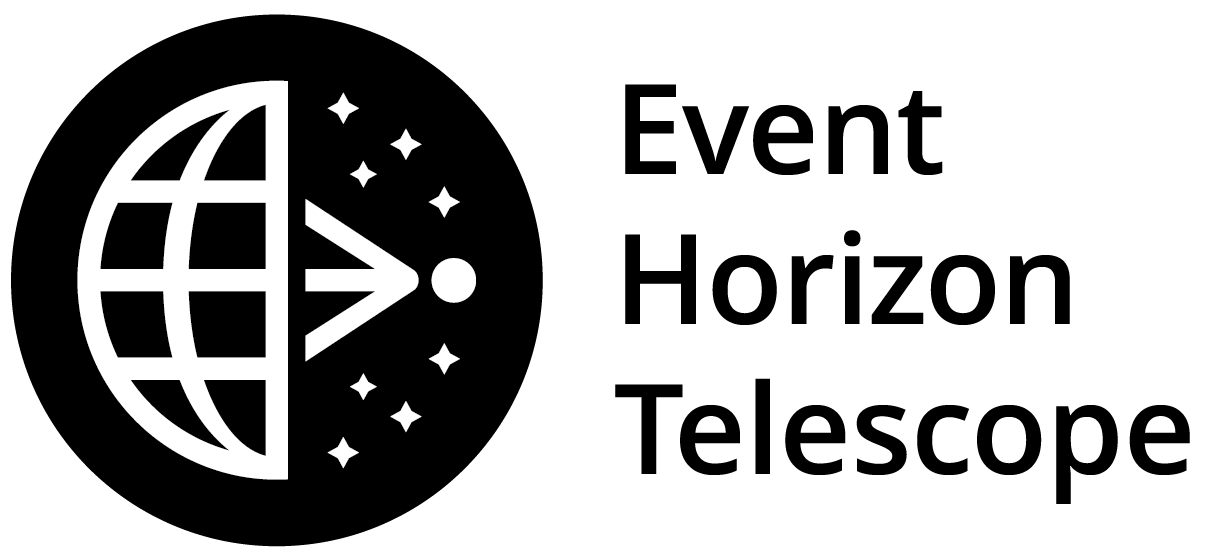}}
    \label{fig:my_label}
\end{minipage}
\begin{minipage}[t]{0.49\textwidth}
\begin{flushright} \large
\textbf{Event Horizon Telescope \\ Memo Series} 
\end{flushright}
\end{minipage}
\end{figure}

\begin{center}
EHT Memo \memonr{}
\end{center}

% Title section: Use the \author, \title, and \date info
{\let\newpage\relax\maketitle}  % Prevent maketitle inserting newpage if possible
\thispagestyle{empty}		% Inhibit the page number on this page

% Add Affiliations
\begin{flushleft}
{\small
\affiliations{}
}
\end{flushleft}

% Add optional History. Make this a compact list
\ifshowhistory{
\setitemize{noitemsep,topsep=0pt,parsep=0pt,partopsep=0pt}  
\begin{flushleft}
\vspace{0.25cm}
Memo history:
\begin{itemize}
\itemsep0em
\memohistory{}
\end{itemize}
\end{flushleft}
}\fi

%%%%%%%%%%%%%%%%%%%%%%%%%%%%%%%%
%%%%%  Optional Abstract   %%%%% 
%%%%%%%%%%%%%%%%%%%%%%%%%%%%%%%%

% \vspace{1cm}
\newpage

\begin{abstract} 

The Event Horizon Telescope (EHT) has been conducting science observations annually since 2021 after the initial campaigns in 2017 and 2018. In this memo, we report on the metadata associated to the observations from 2021 April and 2022 March at 1.3\,mm wavelengths. We follow a similar path for the collection, processing, and validation of the metadata used for the initial amplitude calibration of the 2018 EHT data. In this release, we introduce some differences regarding the treatment of sensitivities and gain curves for some stations. In addition, we describe the changes to the EHT array since 2018. 

\end{abstract} 

\vspace{1cm}

%%%%%%%%%%%%%%%%%%%%%%%%%%%%%%%%
%%%%%  Optional TOC        %%%%% 
%%%%%%%%%%%%%%%%%%%%%%%%%%%%%%%%
% \newpage

\tableofcontents

%%%%%%%%%%%%%%%%%%%%%%%%%%%%%
%%%%%  Start Report   %%%%% 
%%%%%%%%%%%%%%%%%%%%%%%%%%%%%

\setlength{\parskip}{\baselineskip}
\parindent 0pt
\vspace{1.0cm}

\newpage

\section{Introduction}\label{sec:introduction}

The Event Horizon Telescope (EHT) is a continually improving array that has been performing yearly science observations since 2021, following the successful 2017 and 2018 campaigns. In this release, we build on the framework set in place in those campaigns for the treatment of the Level 1 (L1) processing of the metadata used for amplitude calibration \citep[][]{issaoun17a, janssen19, koay23a}, highlighting the changes made in the 2021 and 2022 campaigns.

The 2021 observations marked the return of EHT operations after the restrictions imposed by the COVID-19 pandemic. Significant adjustments were needed to enable a new EHT campaign, such as controlling some telescopes remotely. For example, the Greenland Telescope (GLT) operations required coordination of efforts between the GLT staff in Taiwan and engineers employed by the Pituffik Space Base in Greenland \citep{chen23}, who had no previous experience with telescope operations. In addition, both the NOrthern Extended Millimeter Array (NOEMA) operated by the Institut de Radio Astronomie Millim\'{e}trique (IRAM) in France, and the University of Arizona 12-meter Telescope at Kitt Peak (KT) joined the EHT observations for the first time in that year. The Large Millimeter Telescope Alfonso Serrano (LMT) in Mexico returned to operations on time for the 2022 campaign, although relying on a crystal clock rather than a maser, which was unavailable due to extended maintenance. Thus, a total of 10 and 11 stations participated in the 2021 and 2022 EHT campaigns, respectively. This makes a total of eight single-dish stations (APEX, \pv{}, SPT, LMT, SMT, JCMT, GLT and KT) and three phased arrays (ALMA, SMA and NOEMA). 

As in 2018, the 2021 and 2022 observations used full polarization in four 2\,GHz-wide frequency bands, covering 8.0\,GHz in the 230\,GHz window. However, the useful bandwidth in each band is slightly reduced by approximately 128\,MHz for ALMA and by 64\,MHz for NOEMA owing to different sampling rates compared to the rest of the EHT stations.

The frequency bands are centered at 213.1 (b1) and 215.1 (b2) for the lower sideband (LSB), and at 227.1 (b3) and 229.1\,GHz (b4) for the upper sideband (USB). Technical observations in the 345\,GHz window were made in 2021 for network validation purposes and are discussed in \citet{goddi25}.

After the 2018 observations, the JCMT switched to a dual-polarization, sideband-separating (2SB) receiver (see Section \ref{sec:gains}). Hence, in the 2021 and 2022 EHT array, only SMA is operating with a double-sideband (DSB) receiver, while the rest of the stations use 2SB receivers. 

As in previous campaigns, ALMA observed in linear polarization and all other sites in circular polarization (RCP and LCP). Hence, a polarization conversion for the ALMA data was needed after correlation for the 2021 and 2022 science data \citep[see][]{martividal16,koay23a}. We note that APEX observed in linear polarization on the first day of observations in 2021, and then circular polarization the rest of the time (see Section \ref{sec:AX}).

In \citet{koay23a} we described the collection, processing, and validation of the 2018 metadata. For subsequent EHT campaigns, we have been preparing consistent guidelines for metadata submission. This eases the processing of the metadata with an ad hoc pipeline, and the production of absolute flux-density calibration tables (in ANTAB\footnote{\url{http://www.aips.nrao.edu/cgi-bin/ZXHLP2.PL?ANTAB}} format). 

In this memo, we present some updates to the metadata validation and processing pipeline that we used in 2018 \citep[sections 5 and 6 in][]{koay23a}, as well as a description of the metadata products in this L1 release for the 2021 and 2022 EHT campaigns. In Section~\ref{sec:gains} we provide estimates of the gains and the gain-elevation curves for each station. In Section~\ref{sec:station_meta} we describe the characteristics of the metadata provided by each station and the factors considered to obtain their sensitivities and the data included in the ANTAB tables. In Section~\ref{sec:pip_update} we describe the updates made to the metadata pipeline. In Section~\ref{sec:l1products} we detail the contents of this L1 release, and the new features with respect to the 2018 release. In addition, Section \ref{sec:pol} summarizes relevant information needed for the polarization calibration of EHT data. In Appendix~\ref{ap:notation} we list the formulae and nomenclature that we use throughout this memo for the flux calibration of the data. Following \citet{koay23a}, in Appendix~\ref{ap:ids} we provide details of the 2021 and 2022 observations to help identify observing schedules (i.e., tracks) and participating sites by referencing codes.

\section{Sensitivities for a-priori calibration}\label{sec:gains}

The antenna sensitivities needed for the ANTAB tables take into account the changes in effective system temperature ($T^{*}_{\rm sys}$; Eq.~\ref{eq:teff}) with elevation. Such changes are written in terms of a maximum gain and a gain curve. The maximum gain is assumed to be constant during the campaign for a given station, frequency band, and polarization. It is given by the so-called degrees per flux unit (DPFU; Eqs.~\ref{eq:dpfusd}, \ref{eq:dpfuph}). The gain curve ($g_E$) is normally expressed by a second-order polynomial of elevation ($E$):

\begin{equation}\label{eq:gc_poly}
g_E = a_0 + a_{1}E + a_{2}E^2,
\end{equation}

where the coefficients $a_0$, $a_1$ and $a_2$ are obtained from the fit to the gain or aperture efficiency values. An alternative representation with coefficients $B$ and $E_0$, directly shows the elevation at which the gain curve peaks ($E_0$):

\begin{equation}\label{eq:gc_alt}
g_E = 1 - B(E -E_0)^2
\end{equation}

\citep[][]{janssen19}. This is a normalized gain curve, so $g_E = 1$ when $E=E_0$. In either form, the gain curve is meant to track differences in gain owing to the structural effects on the dish when observing at different elevations. 

Each station provides the information needed to obtain their characteristic sensitivity, or directly provides the DPFUs and gain curve coefficients (see details in Section~\ref{sec:station_meta}). These quantities are used to convert the amplitudes to physical units in terms of the system equivalent flux density (SEFD; Eqs.~\ref{eq:sefdsd}, \ref{eq:sefdph}). In Table~\ref{tab:gains} we summarize the DPFU values (per sideband and polarization) for the stations that participated in the 2021 and 2022 campaigns, as well as the parameters $B$ and $E_0$ that define the gain curve, noting that when these parameters are zeros, it is because the gain curve is assumed to be flat, and thus $g_E = 1$.

\begin{table*}[t]
\caption{DPFUs and gain curve parameters for the EHT stations in the 2021 and 2022 campaigns. DPFU values are given per sideband and polarization; these represent combined values from individual antennas. When given for the phased arrays, DPFUs include a factor for the number of antennas ($N_{\rm ant}$) forming the phased array.}\label{tab:gains} 
\vspace{-0.25cm}

\begin{tabular*}{\linewidth}{@{\extracolsep{\fill}}l c c c c c c c}
\toprule 
\toprule
\multirow{2}{*}[-1pt]{Station} &  \multicolumn{5}{c}{DPFU\,(K\,Jy$^{-1}$) in the 2021 campaign} & \multicolumn{2}{c}{$g_E$} \\ 
      \cmidrule(lr){2-6}  
      \cmidrule(lr){7-8}  
& LSB-RCP & LSB-LCP & USB-RCP & USB-LCP & uncertainties & $B$ & $E_0$\\ 
\hline  
 ALMA & 0.82 & 0.82 & 0.82 & 0.82 & $\pm$\,10\% & 0 & 0 \\
 APEX & 0.0268 & 0.0278 & 0.0272 & 0.0280 & $\pm$\,(5.0 -- 5.2)\,\% & 0 & 0 \\
 GLT & 0.0275 &  0.0265 & 0.0275 & 0.0265 & $\pm$\,(5.3 -- 5.9)\,\% & 0 & 0 \\
 \kt{} & 0.0230 & 0.0231 & 0.0223 & 0.0231 & $\pm$\,(8.9 -- 9.1)\,\% & 7.3$\times10^{-5} \pm$ 7.9\,\%  & 43.2 $\pm$ 6.2\,\% \\ 
 SMT & 0.0177 & 0.0166 & 0.0170 & 0.0160 & $\pm$\,(5.8 -- 6.0)\,\% & 8.2$\times10^{-5} \pm$ 10.4\,\% & 57.6 $\pm$ 2.0\%\\
 JCMT & 0.0330 & 0.0330 & 0.0330 & 0.0330 & $\pm$\,13.2\,\% & 0 & 0\\
 NOEMA & 0.3011 & 0.3044 & 0.2994 & 0.3036 & $\pm$\,10\,\% & 0 & 0\\
 \pv{} & 0.1300 & 0.1258 & 0.1059 & 0.1232 & $\pm$\,(15.8 -- 17.8)\% &  1.71$\times10^{-4} \pm$ \,4.4\%  &  44.1 $\pm$ 1.2\,\% \\
 SMA & 0.0298 & 0.0345 & 0.0298 & 0.0345 & $\pm$\,(5 -- 15)\% & 0 & 0\\
 SPT & 0.0068 & 0.0071 & 0.0068 & 0.0071 & $\pm$\,(6.0 -- 7.2)\% & 0 & 0\\
\midrule
\multirow{2}{*}[-1pt]{Station} &  \multicolumn{5}{c}{DPFU\,(K\,Jy$^{-1}$) in the 2022 campaign} & \multicolumn{2}{c}{$g_E$} \\ 
      \cmidrule(lr){2-6}  
      \cmidrule(lr){7-8}  
& LSB-RCP & LSB-LCP & USB-RCP & USB-LCP & uncertainties & $B$ & $E_0$\\ 
\hline  
 ALMA & 1.18 & 1.18 & 1.18 & 1.18 & $\pm$\,10\% & 0 & 0 \\
 APEX & 0.0280 & 0.0288 & 0.0278 & 0.0283 & $\pm$\,(5.1 -- 5.5)\,\% & 0 & 0 \\
 GLT & 0.0270 &  0.0269 & 0.0270 & 0.0269 & $\pm$\,6\,\% & 0 & 0 \\
 \kt{} & 0.0225 & 0.0227 & 0.0225 & 0.0227 & $\pm$\,(12.2 -- 12.3)\,\% & 7.3$\times10^{-5} \pm$ 7.9\,\%  & 43.2 $\pm$ 6.2\,\% \\ 
 LMT & 0.1729 & 0.1761 & 0.1729 & 0.1761 & $\pm$\,(23.9 -- 24.1)\% & 0 & 0\\
 SMT & 0.0185 & 0.0174 & 0.0179 & 0.0168 & $\pm$\,(6.0 -- 6.1)\,\% & 8.2$\times10^{-5} \pm$ 10.4\,\% & 57.6 $\pm$ 2.0\%\\
 JCMT & 0.0330 & 0.0330 & 0.0330 & 0.0330 & $\pm$\,13.2\,\% & 0 & 0\\
 NOEMA & 0.3428 & 0.3460 & 0.3277 & 0.3284 & $\pm$\,10\,\% & 0 & 0\\ 
 \pv{} & 0.1356 & 0.1324 & 0.1310 & 0.1280 & $\pm$\,(14.5 -- 15.9)\% &  1.71$\times10^{-4} \pm$ \,4.4\%  &  44.1 $\pm$ 1.2\,\% \\
 SMA & 0.0435 & 0.0274 & 0.0349 & 0.0277 & $\pm$\,(5 -- 15)\% & 0 & 0\\
 SPT & 0.0087 & 0.0078 & 0.0087 & 0.0078 & $\pm$\,(15.4 -- 23.7)\% & 0 & 0\\
\bottomrule
\end{tabular*}
\newline
\end{table*}

The DPFU values in Table~\ref{tab:gains} are mean values for the lower and upper sidebands in the corresponding observing campaign. The header of the ANTAB tables used for a-priori calibration contains band-specific DPFUs for b1, b2, b3, and b4, as well as gain curve coefficients presented in the form of a second-order polynomial function of the elevation. For phased arrays, there are further differences explained in Sections~\ref{sec:AA}, \ref{sec:NN}, and \ref{sec:SW}.

As a general rule to calculate the uncertainty of each mean DPFU value, we use the standard deviation in the {\fontfamily{qcr}\selectfont python} library {\fontfamily{qcr}\selectfont pandas}, which applies the Bessel's correction by default. DPFU errors also include in quadrature the uncertainty for the planet's flux density models if these were used to obtain aperture efficiencies. A 10\,\% uncertainty is assumed when using Saturn (owing to its rings), and 5\,\% is considered for other planets \citep[][Bryan Butler and Dan Marrone, priv.\,comm.]{koay23a}. The LMT observed a line source to estimate the aperture efficiency, and their nominal flux calibration uncertainty is $\sim10$\,\%, which we add in quadrature. 

In Table \ref{tab:multiyr_sensitivities} we provide representative sensitivities measured for scans of the primary targets, M87$^*$ and Sgr\,A$^*$. We include the sensitivities reported for the 2017 \citep{EHTC2019c} and 2018 \citep{koay23a} campaigns for comparison. 

\begin{table*}[t]
\caption{Representative sensitivities for primary targets in the EHT campaigns from 2017 to 2022. Values from 2017 and 2018 come from \citet{EHTC2019c} and \citet{koay23a}, respectively. Aperture efficiencies ($\eta_{\rm A}$) and DPFUs are averaged mean values over frequency bands and polarizations.}\label{tab:multiyr_sensitivities}
\vspace{-0.25cm}
\begin{tabular*}{\linewidth}{@{\extracolsep{\fill}}l c c c c  c c c c}
\toprule 
\toprule
\multirow{2}{*}[-1pt]{Station} & \multicolumn{4}{c}{$\eta_{\rm A}$\,(\%)\textsuperscript{a}} & \multicolumn{4}{c}{DPFU\,(K\,Jy$^{-1}$)\textsuperscript{b}} \\ 
      \cmidrule(lr){2-5}  
      \cmidrule(lr){6-9}  
~  & 2017 & 2018 & 2021 & 2022 & 2017 & 2018 & 2021 & 2022 \\
\hline  
ALMA  	&  68 & 68 & 68  &  68  & 1.0300\,(37) & 1.1900\,(43) & 0.8200\,(29) & 1.1800\,(42) \\
APEX 	&  61 & 64 & 67  &  69  & 0.0250       & 0.0260       & 0.0275       & 0.0282       \\ 
GLT 	&  -  & 22 & 66  &  66  & -            & 0.0088       & 0.0270       & 0.0269       \\ 
\kt{}   &  -  & -  & 56  &  55  & -            & -            & 0.0229       & 0.0226       \\ 
LMT 	&  28 & 22 & -   &  24  & 0.0830       & 0.1570       & -            & 0.1745       \\ 
SMT 	&  60 & 65 & 59  &  62  & 0.0170       & 0.0190       & 0.0168       & 0.0176       \\ 
JCMT 	&  52 & 46 & 52  &  52  & 0.0330       & 0.0300       & 0.0330       & 0.0330       \\ 
NOEMA 	&   - & -  & 55  &  55  & -            & -            & 0.3021\,(9)  & 0.3362\,(10) \\ 
\pv{} 	&  47 & 55 & 47  &  51  & 0.0340\textsuperscript{c}       & 0.1400       & 0.1212       & 0.1318       \\ 
SMA  	&  75 & 75 & 75  &  75  & 0.0460\,(6)  & 0.0400\,(7)  & 0.0322\,(6)  & 0.0334\,(6)  \\ 
SPT 	&  60 & 25 & 25  &  29  & 0.0061       & 0.0071       & 0.0070       & 0.0083       \\
\midrule
\multirow{2}{*}[-1pt]{Station} &  \multicolumn{4}{c}{Median $T^*_{\rm sys}$\,(K)} & \multicolumn{4}{c}{Median SEFD\,(Jy)} \\ 
      \cmidrule(lr){2-5}  
      \cmidrule(lr){6-9}  
~  & 2017 & 2018 & 2021 & 2022 & 2017 & 2018 & 2021 & 2022 \\
\hline
ALMA  	&   76 & 113 & 82  & 88  &  74     & 106    & 113    & 84     \\
APEX 	&  118 & 163 & 147 & 122 &  4,800  & 6,259  & 5,350  & 4,321  \\ 
GLT 	&  -   & 410 & 576 & 326 &  -      & 46,287 & 21,183 & 12,117 \\ 
\kt{}   &  -   & -   & 354 & 209 &  -      & -      & 15,478 & 9,207  \\ 
LMT 	&  371 & 208 & -   & 186 &  4,500  & 1,362  & -      & 1,068  \\ 
SMT 	&  291 & 375 & 376 & 189 &  17,100 & 20,330 & 21,607 & 10,703 \\ 
JCMT 	&  345 & 379 & 154 & 185 &  10,500 & 12,816 & 4,664  & 5,592  \\ 
NOEMA 	&  -   & -   & 303 & 373 &  -      & -      & 1,021  & 1,113  \\ 
\pv{} 	&  226 & 351 & 357 & 228 &  6,900\textsuperscript{c}  & 2,539  & 2,961  & 1,746  \\ 
SMA  	&  285 & 283 & 229 & 338 &  6,400  & 8,458  & 7,121  & 10,136 \\ 
SPT 	&  118 & 95  & 92  & 109 &  19,300 & 13,365 & 13,237 & 13,034 \\
\bottomrule
\end{tabular*}
\newline
\vspace{0.05cm}
\textsuperscript{a}\footnotesize{The aperture efficiency for phased arrays corresponds to the nominal efficiency of a single dish element.}\\
\textsuperscript{b}\footnotesize{The number in parenthesis in the listed values indicates the number of phased antennas ($N_{\rm ant}$) considered when obtaining the DPFU for phased arrays.}\\
\textsuperscript{c}\footnotesize{This lower DPFU and higher SEFD in 2017 arose from a technical issue with the maser and does not reflect the actual telescope surface accuracy \citep[see][]{EHTC2019c}.}
\end{table*}

Comparing 2021--2022 values with those reported from 2017--2018, we note the following:
\vspace{-0.35cm}
\begin{itemize}
    \item The GLT improved its sensitivity for the 2021 EHT campaign, since the antenna was fully commissioned for science observations by then \citep[][]{koay23b}. In addition, the GLT VLBI backends were upgraded in 2019, and observations at both LSB and USB could be performed.
    \item The sensitivity of the SMA appears to be slightly worse in 2021--2022. However, the change we see in the DPFUs is related to the way in which values were obtained. In 2018, the reported value corresponds to the median instead of the average we report here. The mean in 2018 is $0.0386$\,K\,Jy$^{-1}$ considering seven phased antennas. Using six antennas as is the case in 2021--2022, the mean DPFU becomes $0.0331$\,K\,Jy$^{-1}$, which is similar to what we report for 2021--2022.
    \item In 2021--2022 the JCMT started using the GLT spare receiver N$\overline{\rm{a}}$makanui \citep{mizuno20, mizuno21}. This upgrade from a DSB to a 2SB receiver is reflected in lower $T^*_{\rm sys}$ values.
\end{itemize}
Other apparent differences between 2017--2018 and 2021--2022, are related to weather conditions and/or technical issues pertaining to each campaign. As a reference to the weather conditions, we show a list of zenith opacities on primary target scans during the 2021--2022 EHT campaigns in Table \ref{tab:tau}.

\begin{table*}[t]
\caption{Median zenith sky opacities (1.3\,mm) per day at each station for primary target scans during the 2021--2022 EHT campaigns.}\label{tab:tau}
\vspace{-0.25cm}
\begin{tabular*}{\linewidth}{@{\extracolsep{\fill}}l c c c c c c c c c c}
\toprule 
\toprule
\multirow{2}{*}[-1pt]{Date} & \multicolumn{10}{c}{Station} \\ 
      \cmidrule(lr){2-11}  
  & ALMA & APEX & GLT & \kt{} & LMT & SMT & JCMT & NOEMA & \pv{} & SPT \\
\hline  
Apr/09/2021 & - & 0.03 & 0.35 & 0.30 & - & 0.15 & 0.05 & 0.15 & - & 0.05 \\
Apr/13/2021 & 0.17 & 0.14 & 0.26 & 0.32 & - & 0.19 & 0.06 & - & 0.36 & - \\
Apr/14/2021 & 0.07 & 0.06 & 0.23 & 0.25 & - & 0.31 & 0.08 & 0.08 & 0.64 & 0.06 \\
Apr/15/2021 & 0.04 & 0.04 & - & 0.25 & - & 0.37 & 0.08 & - & - & 0.05 \\
Apr/16/2021 & - & - & - & 0.30 & - & 0.23 & 0.06 & - & - & 0.06 \\
Apr/17/2021 & 0.05 & 0.05 & 0.25 & 0.23 & - & 0.20 & 0.07 & 0.25 & 0.20 & 0.04 \\
Apr/18/2021 & 0.05 & 0.04 & 0.41 & 0.23 & - & 0.15 & 0.08 & 0.19 & 0.12 & - \\
\midrule
Mar/18/2022 & - & - & 0.17 & - & 0.15 & 0.18 & 0.12 & 0.42 & 0.30 & - \\
Mar/19/2022 & - & 0.04 & 0.19 & - & - & 0.11 & - & 0.19 & 0.21 & - \\
Mar/20/2022 & 0.09 & 0.04 & - & - & 0.11 & 0.22 & 0.11 & 0.39 & 0.18 & 0.05 \\
Mar/22/2022 & 0.06 & 0.04 & 0.23 & - & 0.12 & 0.20 & 0.09 & 0.07 & 0.15 & 0.04 \\
Mar/23/2022 & 0.04 & 0.04 & - & - & 0.11 & 0.04 & 0.14 & 0.14 & - & 0.05 \\
Mar/27/2022 & 0.05 & 0.04 & 0.22 & - & 0.10 & 0.25 & 0.12 & 0.18 & 0.15 & - \\
\bottomrule
\end{tabular*}
\newline
\vspace{0.05cm}
\footnotesize{Notes: (1) SMA included no opacity values in their metadata, but JCMT could be used as a reference. (2) In 2022, \kt{} did not provide opacity values. (3) Primary target scans are not included in the observations from 26 March 2022, so we do not report opacity values for that day in this Table.}
\end{table*}

\section{Metadata characteristics per station and construction of ANTAB tables}\label{sec:station_meta}

Although each station in the EHT array provides raw metadata in a standard format \citep[section~3 in][]{koay23a}, there are still some peculiarities in each package that need to be considered. In the following subsections, we provide details of the information provided by each station and how this is used to assemble the EHT ANTAB tables. When required, we also describe the different issues that affect the data.

\subsection{ALMA}\label{sec:AA}

The phased-ALMA calibration workflow is described in detail in \citet{goddi19}. The Level 2 Quality Assurance (QA2) process carried out by the Joint ALMA Observatory produces calibration tables needed to perform the polarization conversion of phased-ALMA visibilities from linear to circular basis using the post-correlation {\fontfamily{qcr}\selectfont PolConvert} software \citep{martividal16, matthews18}.

{\fontfamily{qcr}\selectfont PolConvert} computes amplitude corrections, which are provided by the QA2 team in ANTAB format for each track and frequency band, in each of the 32 spectral windows (each with 56\,MHz bandwidth) at a rate of $\sim 0.4$\,s. These corrections are stored as phased-array $T^{*}_{\rm sys}$ values %per integration time 
valid for both RCP and LCP, and with a constant antenna-wise averaged DPFU value, which has been set to 0.031\,K\,Jy$^{-1}$ at 230\,GHz (Band 6) since the EHT campaign in 2017 \citep{goddi19}. These $T^{*}_{\rm sys}$ values already account for elevation- and time-dependent gains calculated through self-calibration (so a flat gain curve can be assumed), as well as for the number of antennas in the phased array and the corresponding phasing efficiencies \citep{goddi19}. As described in \citet{koay23a}, we perform data averaging every 6\,s in each spectral window to avoid problems with calibration tools which cannot handle the high cadence at which $T^{*}_{\rm sys}$ values are provided. These time-averaged $T^{*}_{\rm sys}$ post-{\fontfamily{qcr}\selectfont PolConvert} values and the fixed DPFU are included in the EHT ANTAB tables.

As part of the metadata, the ALMA team also provides phasing efficiencies and measured $T^{*}_{\rm sys}$ values obtained before {\fontfamily{qcr}\selectfont PolConvert} (which do not account for the number of antennas in the array or the phasing efficiencies). Both of these quantities are provided per track, frequency band, polarization component, and integration time. The median phasing-efficiency values achieved in 2021 and 2022 for scans on primary targets are 93\,\%, and 98\,\%, respectively. In 2021, an average of 29 antennas observed during the EHT observations, resulting in an equivalent phased-array diameter of $\sim65$\,m. In 2022 the average number of participating antennas was 42, resulting in an equivalent diameter of $\sim78$\,m.

To calculate characteristic ALMA SEFDs, we can apply a theoretical approach which requires using Eq. \ref{eq:sefd_th} \citep{goddi19} with $\eta_{\rm A} = 0.68$ \citep[][Table~\ref{tab:multiyr_sensitivities}]{matthews18}, the pre-{\fontfamily{qcr}\selectfont PolConvert}  $T^{*}_{\rm sys}$ values, and the DPFU in Table \ref{tab:gains} (which does not include the phasing efficiency term). Alternatively, we may use the DPFU and post-{\fontfamily{qcr}\selectfont PolConvert}  $T^{*}_{\rm sys}$ values estimated in the QA2 gain calibration which are included in the ANTAB tables provided by ALMA. The SEFD medians estimated with these two methods result in similar values \citep{goddi19, crew23}. For convenience, we use the theoretical approach, since pre-{\fontfamily{qcr}\selectfont PolConvert}  $T^{*}_{\rm sys}$ values are identified not only by a timestamp (as in the QA2 ALMA ANTAB tables) but also by source, so we can easily separate the values corresponding to the primary targets. 

\begin{equation}\label{eq:sefd_th}
{\rm SEFD}_{\rm th} = \frac{\langle{T^{*}_{\rm sys}\rangle \,2k_{\rm B}}}{\eta_{\rm A}\, A_{\rm geom}}
\end{equation}

The $A_{\rm geom}$ term in Eq.~\ref{eq:sefd_th} is the total collecting area of the array, so it includes the number of phased antennas. In the ALMA Phasing System, the number of phased antennas is odd by construction \citep{matthews18}. But in Table \ref{tab:multiyr_sensitivities} we present average values which in principle could also be even, as it happens in 2022.

\subsection{APEX}\label{sec:AX}

The APEX raw metadata packages include documents that describe how APEX gains were obtained\footnote{We note that different naming conventions for polarization products are used in these documents. Polarization 1 and X correspond to LCP, whilst 2 and Y correspond to RCP.}. The measurements are based on the observation of planets: Mars and Jupiter in 2021, and Mars and Venus in 2022. In 2021, the team obtained DPFU factors by beam-correcting the planets' flux densities obtained from the {\fontfamily{qcr}\selectfont ASTRO} program of the IRAM {\fontfamily{qcr}\selectfont GILDAS}\footnote{\url{https://www.iram.fr/IRAMFR/GILDAS}} software, as well as from the Common Astronomy Software Application ({\fontfamily{qcr}\selectfont CASA}) package \citep{mcmullin07,butler12}. In 2022 only {\fontfamily{qcr}\selectfont CASA} was used. No significant elevation dependence was found; thus we have used a flat gain curve in the header of the 2021--2022 ANTAB tables. 

In 2021, some technical issues affected the ANTAB tables and the science data \citep[section B.3. in][]{EHTC2025b}. All scans from track e21b09, and scans No0012 to No0034 from track e21e13 were observed with wobbler switching, using a duty cycle of 50\%. Thus, in this release we have applied a correction factor of two in the $T^*_{\rm sys}$ values for the affected scans. In addition, during track e21b09 the quarter-wave plate (QWP) was off by 45\degr{} and as a result the whole track was observed in linear polarization. This issue could be corrected in the science data without the need of {\fontfamily{qcr}\selectfont PolConvert} \citep{EHTC2025b}. No conversion to circular basis is available for the $T^*_{\rm sys}$ values, and thus the header of the APEX $T^*_{\rm sys}$ section in the corresponding ANTAB tables indicate that the polarization is linear. To avoid mixing linear with circular polarization, we exclude the measurements from track e21b09 when obtaining the $T^*_{\rm sys}$ and SEFD median values reported in Table \ref{tab:multiyr_sensitivities}.  

\subsection{GLT}\label{sec:GL}

The GLT underwent a series of adjustments  that significantly improved its surface and pointing accuracy compared to its first participation in the EHT observations in 2018. Also, starting in 2019, the VLBI backends at the GLT allow a recording bandwidth of 64\,Gbps, which is a factor of two better than in 2018 \citep{chen23}. The antenna status during the 2021 EHT campaign is fully described by \citet{koay23b}.

Bad weather and receiver issues prevented the GLT from obtaining efficiency measurements for the 2021 campaign. In 2022, telescope operations were also affected by bad weather. There is a gap in efficiency measurements from March 2020 until March 2023, and no other adjustments to the GLT could be made in this time period.

The efficiency measurements in 2021 are based on Venus observations from March 2020. For 2022, we use a combination of observations on Venus and Jupiter performed in March 2020 and March 2023, which seem to be consistent. These measurements show an improvement by a factor of three over the 2018 values (Table \ref{tab:gains}). Since the GLT is located at a high latitude, the VLBI targets do not change significantly in elevation during the EHT observations; hence we assume a flat gain curve in 2021--2022 as done in 2018.

While the GLT now observes in both LSB and USB, $T^*_{\rm sys}$ values for both RCP and LCP originate from measurements in USB bands b3 \& b4 only, but are applied to all four bands. We note that the GLT did not provide $T^*_{\rm sys}$ values for track e21e13 owing to power-meter issues. We thus resort to obtain modeled values using the prescription found in section~5.2.4 from \citet{koay23a}. We obtain effective temperature of the atmosphere ($T_{\rm atm}$) and opacity values for 13 April at the GLT location from {\fontfamily{qcr}\selectfont ngehtsim} \citep{pesce24}, which has compiled atmospheric state information from NASA's MERRA-2 database \citep{rienecker11,molod15,gelaro17} processed through the \textit{am} radiative transfer code \citep{paine22}. We assume a value of 282\,K for the ambient temperature in the GLT receiver cabin ($T_{\rm amb}$) from the records on that day, and a typical value of 100\,K for the receiver temperature ($T_{\rm rx}$) in Band 6. However, inspection of the calibrated data revealed that large amplitude gain errors remain \citep{EHTC2025b}, and further adjustments were made in the downstream analysis, which are not reflected in the ANTAB tables of this release.

In 2022, the GLT LO frequency had an offset of 189\,MHz. This does not affect the metadata, but is relevant information for obtaining fringes between the GLT and other telescopes in the EHT array.  

On occasions, the GLT observed under light snow and high wind speed ($\gtrsim10$\,m\,s$^{-1}$) conditions. This affects the pointing and thus the quality of the data, so we expect larger uncertainties in the DPFU values we report in Table~\ref{tab:gains}. In 2021, the median opacity at the GLT site is $\gtrsim30$\,\% larger than in 2022 when considering scans on primary targets. This may explain the lower sensitivities obtained in 2021 compared to 2022.

\subsection{\kt{}}\label{sec:KT}

The \kt{} telescope participated in the EHT array for the first time in 2021. Like the GLT, the telescope is one of the prototype ALMA dishes with a 12\,m diameter and an Altitude--Azimuth (Alt--Az) mount \footnote{\url{https://aro.as.arizona.edu/?q=facilities/uarizona-aro-12-meter-telescope}}. 

DPFUs and gain curve information included in the header of the ANTAB tables was provided by the telescope staff. The DPFUs in 2021 come from measurements on Jupiter and Mars obtained from 9 to 18 April 2021, and those in 2022 were obtained based on observations on Jupiter, Mars and Saturn from 18 to 26 March 2022.

\begin{figure}[h]
    \centering
    \includegraphics[width=0.8\textwidth]{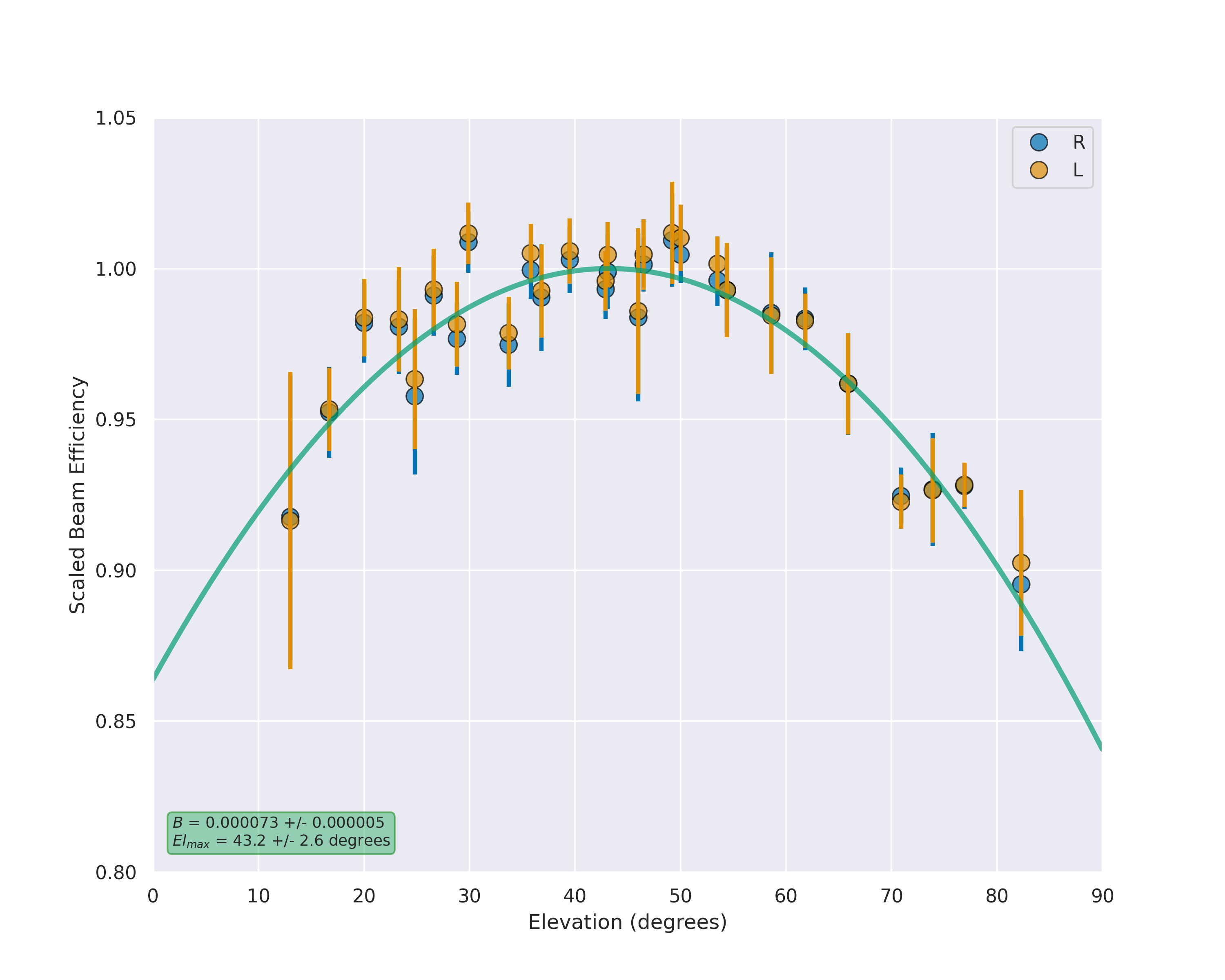}
    \caption{The \kt{} gain curve obtained in 2021. The efficiencies have been normalized by the mean value at the plateau covering the elevation range 35\degr{}--50\degr{}. The solid line represents the fit provided by the telescope representatives.}\label{fig:kt_gc}
    %% Note that the individual uncertainties of the measurements were not taken into account to obtain the fit provided by the telescope representatives  (solid line).
\end{figure}

The 2021 package includes the efficiency measurements\footnote{The efficiency measurement values in their file are given for horizontal and vertical polarizations, which correspond to RCP and LCP, respectively.} that were used by the \kt{} team to fit the gain curve, which is provided in the polynomial representation as $g_E = 0.86395154 + 6.29553363\,{\rm e}{-3}\times~E - 7.283019\,{\rm e}{-5}\times~E^2$. The uncertainties of the gain-curve polynomial coefficients were not provided in the raw metadata package, so we determine them for this release. In Table \ref{tab:gains} we show the gain-curve coefficients in the alternative, normalized representation. Figure \ref{fig:kt_gc} shows the gain curve, which is valid for 2021--2022.

\subsection{LMT}\label{sec:LM}

After the COVID-19 pandemic outbreak, the LMT was able to rejoin the EHT array in 2022. The LMT team provided a table with efficiency measurements obtained on 25 March 2022, based on observations of the star GL2155. We use the mean value of the measurements and consider a 50\,m dish diameter in Eq. \ref{eq:dpfusd} to obtain representative DPFU values at each band and polarization in 2022.

Since 2018, a series of works at the LMT were made to improve its surface, and the change between 2018 and 2022 is deemed substantial. Measurements obtained up to 2023 are not sufficient to adequately characterize the gain curve. Because the LMT has panel actuators to adjust its surface, we have assumed that a flat gain curve is the best approximation. 

The $T^*_{\rm sys}$ tables provided by the LMT team only contain values in the columns corresponding to bands b1 and b2 in RCP and LCP. At this telescope, the $T^*_{\rm sys}$ measurements are made before the band separation at the block downconverters (BDCs). Hence, the $T^*_{\rm sys}$ values in the provided tables have to be relabeled as follows: b1r corresponds to b3l and b4l; b1l corresponds to b1l and b2l; b2r corresponds to b1r and b2r; and b2l corresponds to b3r and b4r. Here, the label ``r'' stands for RCP, and ``l'' for LCP.

Another characteristic of the LMT during the 2022 observations is that its maser was not available and a crystal was used for timing. This needs to be considered in downstream analysis to account for delays in the signal and losses in sensitivity.

\subsection{SMT} \label{sec:MG}

The SMT team provides a dedicated file with DPFU and gain-curve parameters as done for \kt{}. According to these files, the gain curve for 2021--2022 is the same as that of the EHT 2017 and 2018 campaigns \citep{issaoun17b, koay23a}, and is expressed in its polynomial form as $g_E = 0.727119 + 9.47339\,{\rm e}{-3}\times~E - 8.222\,{\rm e}{-5}\times~E^2$. The DPFU values are based on observations to Jupiter and Mars made on 9--18 April 2021, and 17--27 March 2022. To determine the flux scale, the team uses the Butler 2018 planet models (Bryan Butler and Dan Marrone, priv.\,comm.). 

We note that on 14 April 2021 (track e21a14), the reported DPFUs are lower by $\gtrsim30$\,\% than for the rest of the days. A possible explanation for this is that the telescope was relatively defocused owing to high wind speeds experienced on that day.

\subsection{JCMT} \label{sec:MM}

In previous EHT campaigns, we have assumed that the JCMT has a flat gain curve on elevation, although with the gains displaying a night/day dependence (which is a well-known effect at the JCMT). 

To characterize the night/day behavior of the gains in the 2017 campaign, a piece-wise function was used to fit aperture efficiency measurements collected from 2006 to 2017 \citep{issaoun18}, and measurements taken in 2018 were added to improve the fit in the following campaign \citep{koay23a}. 

The piece-wise function used in 2017 and 2018 to fit the night/day behavior of the gains, requires manually setting the time of the day at which the antenna starts and stops being affected by solar irradiance. We expect that these times change throughout the year together with sunrise and sunset times. However many other factors affect these start/stop times, such as the time it takes to cool down different components of the telescope structure, the air temperature around the telescope and thus the time at which the dome was opened before the observations, etc. JCMT staff also notices that the efficiency is subject to a large scatter in daytime owing to the worsening of the seeing which results in pointing shifts and jitter during daytime observations, and is affected as well by the weather and by how close the pointing position is relative to the Sun.

In this release, we adopt a single function to describe the night/day behavior of the gains. An upside-down normal distribution loss function allows to fit all the data without choosing start/stop times when the losses in gain occur \citep[e.g.,][]{drain96}. We fit Eq. \ref{eq:udnlf} to the JCMT gains dependent on time $t_{\rm HST}$, which indicates Hawaii Standard Time:
\begin{equation}\label{eq:udnlf}
\eta_{\rm A}(t_{\rm HST}) = q_0 - q_1 \times {\rm exp}\left(- \frac{\left(t_{\rm HST} - q_2 \right)^2}{2\,q^2_3} \right).
\end{equation}
Here, $q_0$ represents the maximum possible efficiency which remains stable during ``night'' times; $q_1$ is the difference between $q_0$ and the minimum expected efficiency that is reached during ``day'' times; $q_2$ is called the process target, equivalent to the time at which the minimum efficiency occurs; and $q_3$ is the shape parameter of the inverted Gaussian. 

We use all the available data between 2006 and 2021 (in 2022 no new measurements were provided). The 2006--2017 data are not distinguished by polarization, unlike the data obtained after 2018. Thus, we average RCP and LCP measurements per scan for the data obtained after 2018. If there is more than one measurement with the same $t_{\rm HST}$, we keep only the newest one. We also apply a cut in efficiency values to remove outliers below 0.25 and above 0.75 before proceeding with the fit. 

\begin{figure}[ht]
    \centering
    \includegraphics[width=0.49\textwidth]{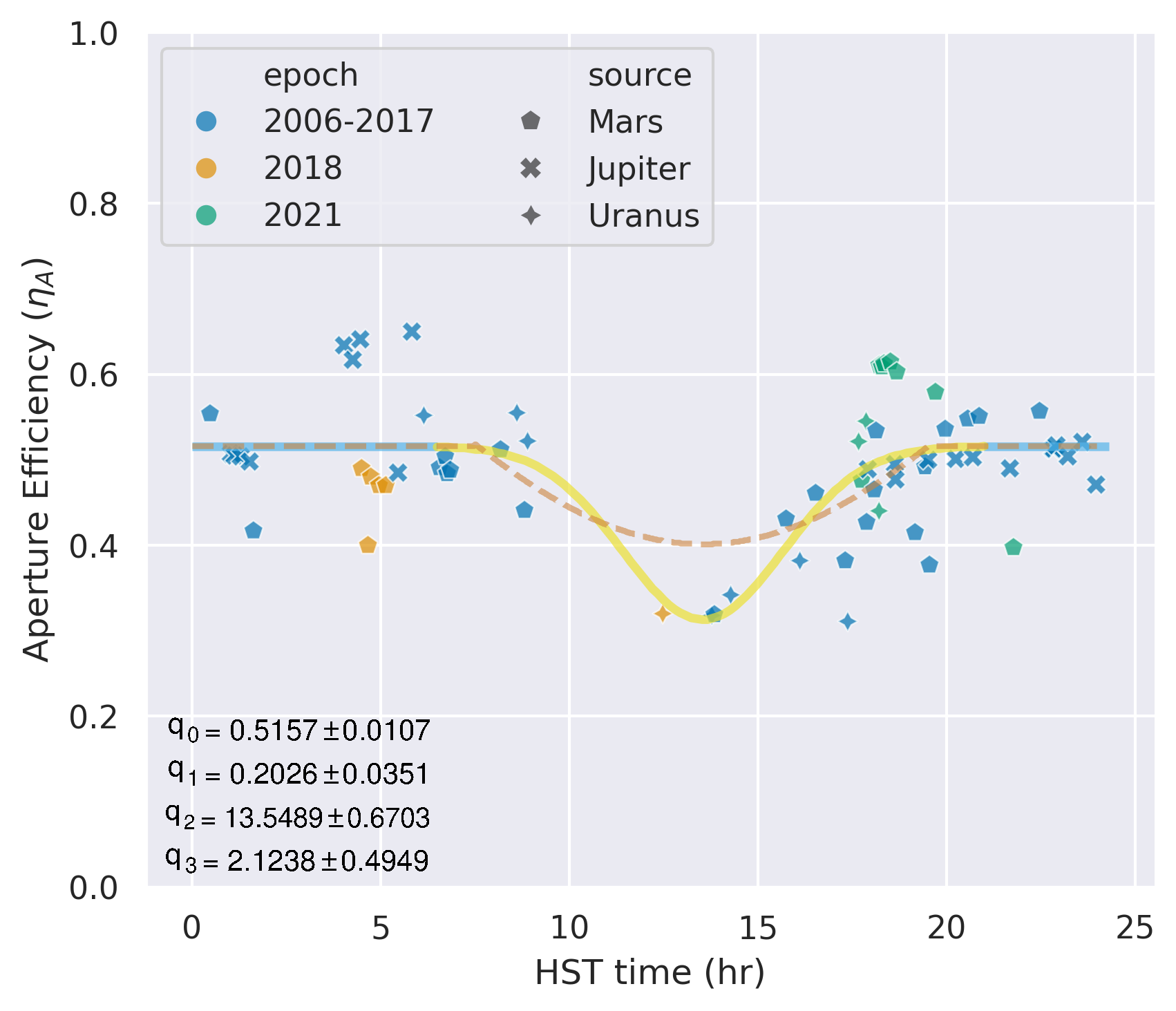}\hfill
    \includegraphics[width=0.49\textwidth]{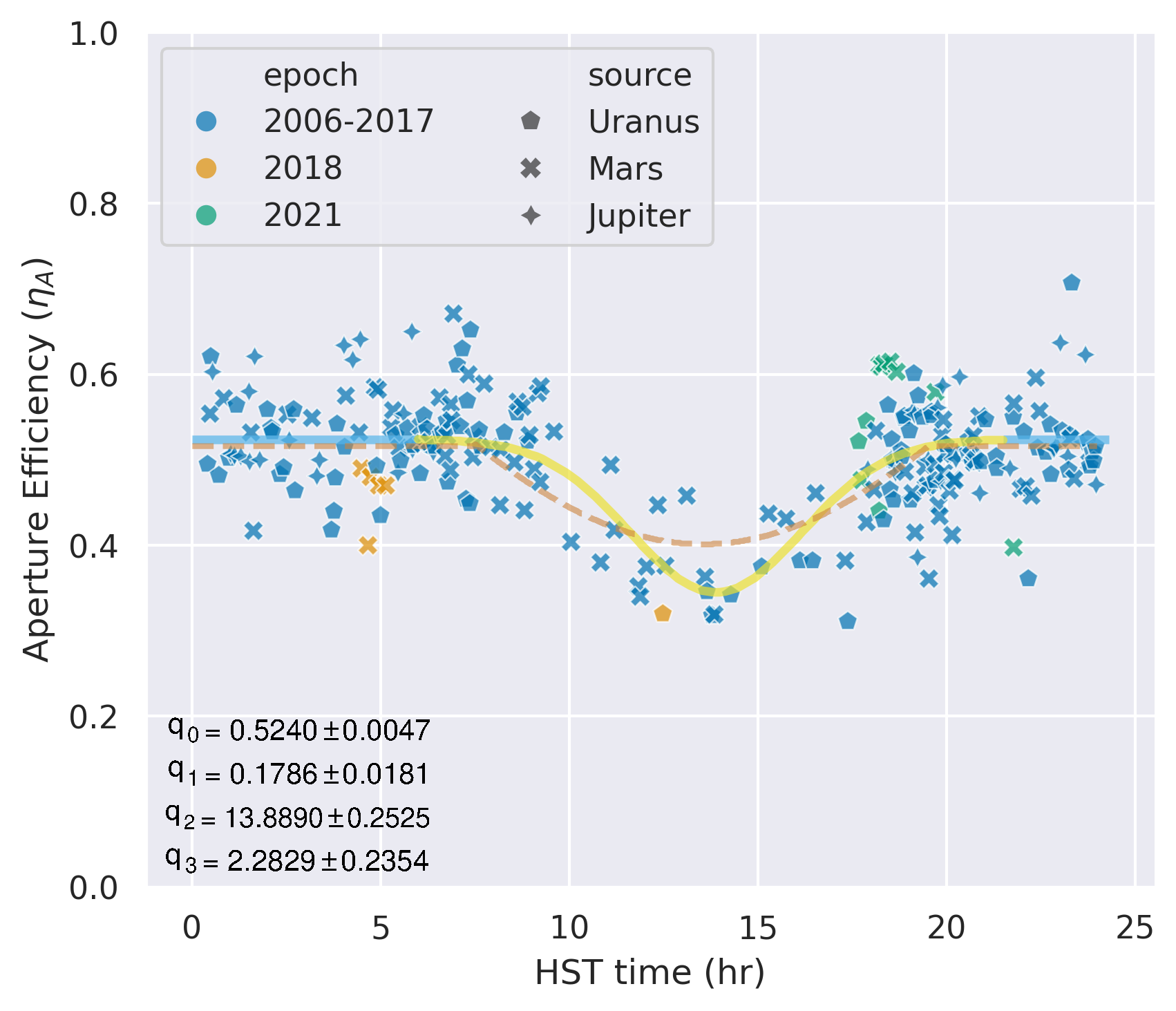}
\caption{JCMT efficiency measurements in typical months of the EHT campaigns (left); all JCMT efficiency data obtained until 2021 (right). In both panels, the dashed, dark-orange line represents the time-dependent gain curve used in 2017; the new fitted curve using Eq. \ref{eq:udnlf} is depicted by the color-coded solid line: the light-blue section indicates the ``night'' times when the gains remain constant, and yellow one indicates the ``day'' times when gain losses occur. The light-blue section lasts from 16:30\,UT to 7:00\,UT (6:30--21:00\,HST) when we fit the data corresponding to March and April, and from 16:00\,UT to 7:30\,UT (6:00--21:30\,HST) when we consider all the data.}\label{fig:jcmt_fit_comp}
\end{figure}

While we do not have a way to control the different factors that influence the night/day behavior of the gains, here we make tests considering only the data taken in typical months of EHT campaigns (March and April), as well as using all the data as done previously for the 2017 fit. In Fig.~\ref{fig:jcmt_fit_comp} we show the resulting fits using Eq. \ref{eq:udnlf} in those two cases. Compared to the 2017--2018 curves, we see that the new fit results in a smoother transition for the night/day behavior of the gains. It also indicates a lower efficiency dip sometime in the middle of the day, when the solar irradiance has a stronger effect on the antenna. We note that the transition between ``night'' and ``day'', differs by half an hour when we compare the two cases considered here.

\begin{figure}[!ht]
    \centering
    \includegraphics[width=0.5\textwidth]{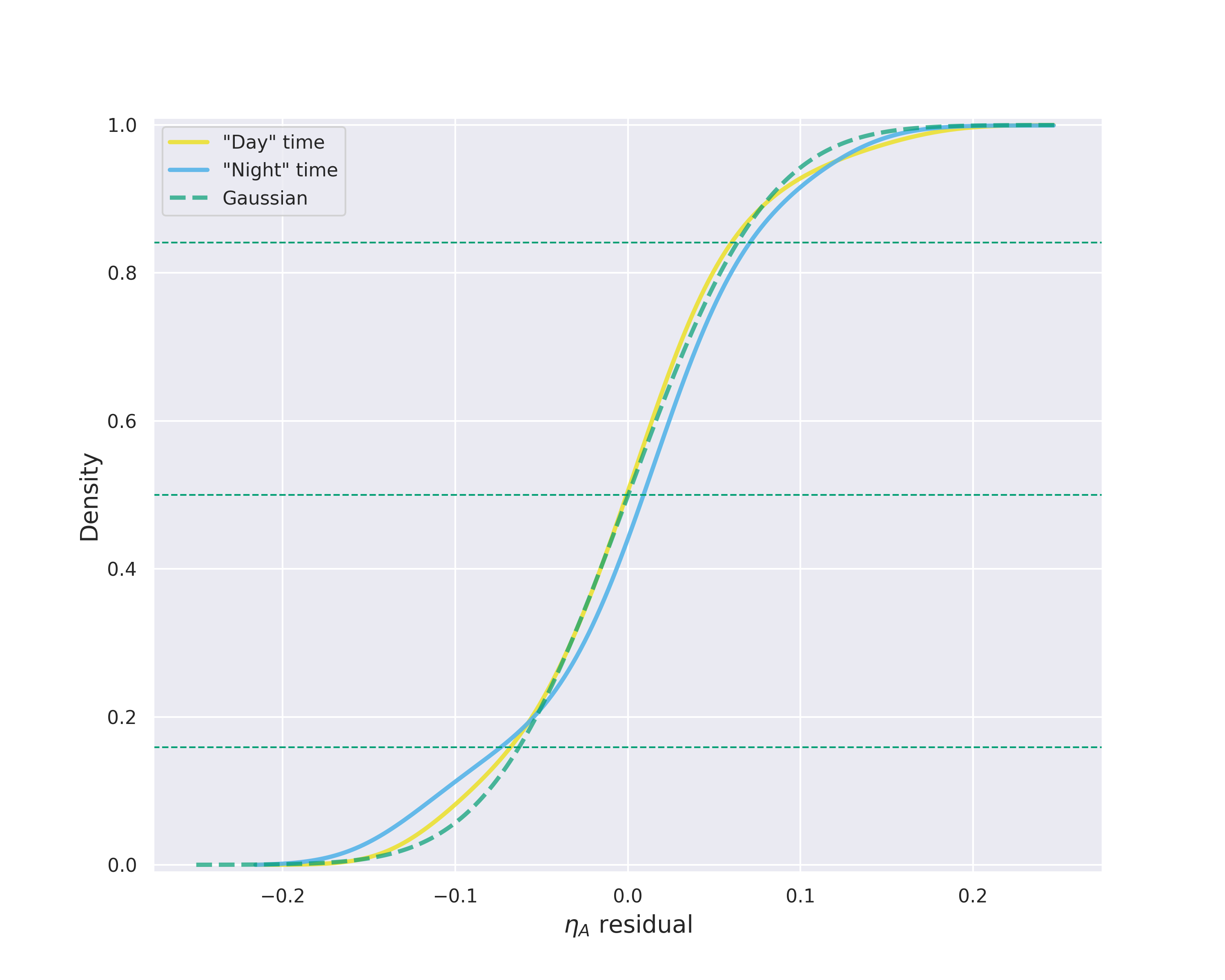}\hfill
    \includegraphics[width=0.5\textwidth]{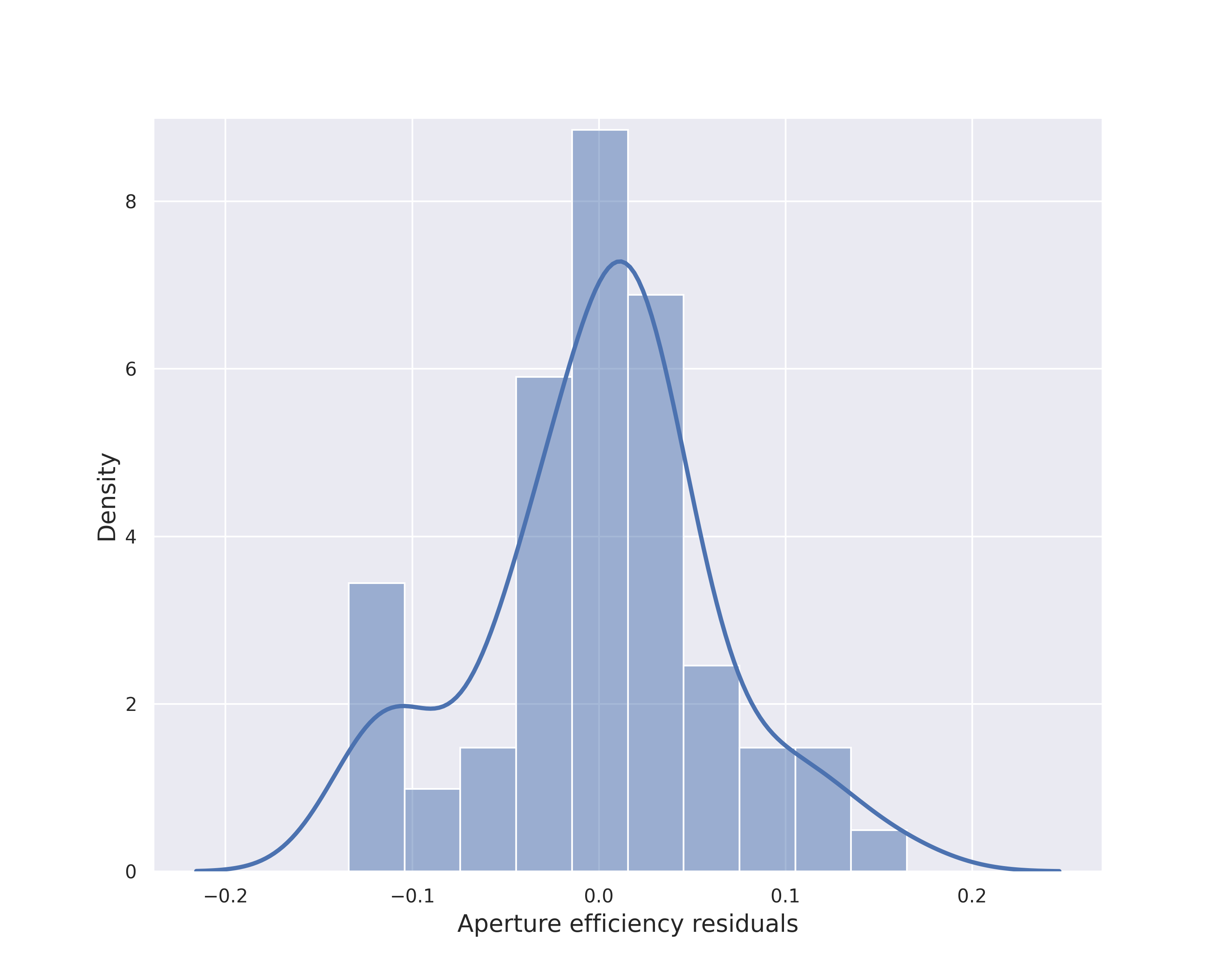}
\caption{JCMT efficiency residuals from measurements taken during the months of March and April. Left: cumulative density plot where ``night'' (solid light-blue line) and ``day'' (solid yellow line) efficiency residuals are shown together with a Gaussian distribution (dashed green line) constructed with the mean of the residuals ($\mu=0.5138$) and their standard deviation ($\sigma=0.0633$). Right: a histogram of the residuals, where the solid blue line represents the curve of the underlying distribution obtained by the {\fontfamily{qcr}\selectfont Seaborn} function  {\fontfamily{qcr}\selectfont histplot}, which uses a kernel density estimate.}\label{fig:jcmt_fit_resid}
\end{figure}

Restricting the data to typical months of the EHT campaign, we find uncertainties in the fitted parameters larger by $\sim10$\,\% than when using all the data. However, we consider this to be the most representative case for our purposes, and we thus use the following parameters for the JCMT time-dependent gain curve: $q_0 = 0.5157 \pm 0.0107$, $q_1 = 0.2026 \pm 0.0351$,  $q_2 = 13.5489 \pm 0.6703$, and  $q_3 = 2.1238 \pm 0.4949$.

To test the goodness of the fit, we use the efficiency residuals for both ``night'' and ``day'' times (Fig. \ref{fig:jcmt_fit_resid}). The residuals roughly follow a Gaussian distribution, albeit with some asymmetries about the mean and a notorious bump at negative values when the total power measurements may have been affected by systematic errors related to e.g., pointing and focus. Using all the residuals, we find a standard deviation of $\sigma=0.0633$, which we use as the representative uncertainty of our fit for both ``night'' and ``day'' times. 

We use Eq.~\ref{eq:dpfusd} with $q_0 = 0.5157$ as the maximum gain and $\sigma=0.0633$ as the representative uncertainty to obtain a DPFU value of $0.0330$\,K\,Jy$^{-1} \pm 13.2$\,\%. This is the value we include in the header of the ANTAB tables in 2021--2022. The time-dependent gain curve normalized by $q_0$:
\begin{equation}\label{eq:jcmt_Norm_timeGainCurve}
g_{\rm c}({t_{\rm HST}}) = 1 - 0.393 \times {\rm exp}\left(- \frac{\left(t_{\rm HST} - 13.549 \right)^2}{2\times2.124^2} \right),
\end{equation}
is used as a correction factor for the JCMT $T^*_{\rm sys}$ values included in the ANTAB tables. %We show this curve in Fig.~\ref{fig:jcmt_Norm_timeGainCurve}.

\subsection{NOEMA} \label{sec:NN}

The NOEMA interferometer on Plateau de Bure, France, joined the EHT science observations in 2021. With up to 12 phased antennas of 15\,m in diameter, this phased array provides a powerful addition to the EHT \citep{EHTC2025b}, equivalent to a 45.5\,m and 47.2\,m diameter antenna in the 2021 and 2022 observations, respectively. The NOEMA phasing system is described in detail by \citet{pietu25}. 

\begin{table*}[th]
\caption{NOEMA bands and their conversion to EHT bands.}\label{tab:noema_eht_bands}
\vspace{-0.25cm}
\begin{tabular*}{\linewidth}{@{\extracolsep{\fill}}c c c  c  c c c}
\toprule 
\toprule
NOEMA & \multirow{2}{*}[-1pt]{Polarization} & Frequency   &~& EHT  & \multirow{2}{*}[-1pt]{Polarization} & Correspondence with \\
Band  & ~                                   & range (MHz) &~& Band &~                                    & NOEMA bands         \\
       \cmidrule(lr){1-3}
       \cmidrule(lr){5-7}
bb1 & RCP  &  209294--213356 &~& b1 &  RCP  & bb1$\times$0.637 $+$ bb2$\times$0.363 \\
bb2 & RCP  &  213356--217418 &~& b1 &  LCP  & bb5$\times$0.637 $+$ bb6$\times$0.363 \\
bb3 & RCP  &  224782--228844 &~& b2 &  RCP  & bb2                               \\
bb4 & RCP  &  228844--232906 &~& b2 &  LCP  & bb6                               \\
bb5 & LCP  &  209294--213356 &~& b3 &  RCP  & bb3                               \\
bb6 & LCP  &  213356--217418 &~& b3 &  LCP  & bb7                               \\
bb7 & LCP  &  224782--228844 &~& b4 &  RCP  & bb3$\times$0.363 $+$ bb4$\times$0.637 \\
bb8 & LCP  &  228844--232906 &~& b4 &  LCP  & bb7$\times$0.363 $+$ bb8$\times$0.637 \\
\bottomrule
\end{tabular*}
\newline
\end{table*}

NOEMA antennas use 2SB receivers\footnote{\url{https://www.iram.fr/IRAMFR/GILDAS/doc/html/noema-intro-html/node5.html}}, which provide an effective bandwidth per sideband of 7.744\,GHz. For EHT observations, the LO frequency is set at 221.1\,GHz. \citet{vonfellenberg25} reports the calibration strategy needed for NOEMA as a station in the 2021 to 2023 EHT campaigns. In Table \ref{tab:noema_eht_bands} we list the characteristics of NOEMA bands, and their correspondence to EHT bands.

The NOEMA team provides metadata packages containing individual antenna gains in Jy\,K$^{-1}$ as well as the following data per observing track: (1) Flag tables; (2) Tables with phasing efficiencies per NOEMA band, at a cadence of 1/sec; (3) Tables with $T^*_{\rm sys}$ values per antenna, per NOEMA band; (4) Tables with phased-array $T^*_{\rm sys}$ values per EHT band and polarization; and (5) Tables with phased-array SEFD values per EHT band and polarization. The phased-array values already incorporate the phasing efficiencies and DPFUs. Because of this, in the ANTAB tables we use a DPFU$=1$ for both polarizations, and the phased-array SEFD values in the NOEMA $T^*_{\rm sys}$ section. The single dishes in NOEMA have no known elevation-dependent gains, so we use a flat gain curve. 

After converting NOEMA bands to EHT bands, we find that the median phasing-efficiency values for scans on primary targets are 92\,\% and 83\,\%, in 2021 and 2022, respectively. 

To obtain the DPFU values in Tables~\ref{tab:gains} and \ref{tab:multiyr_sensitivities}, we take the ratio of phased-$T^*_{\rm sys}$ to phased-SEFDs. The DPFUs in Table~\ref{tab:multiyr_sensitivities} agree within $\lesssim2$\,\% with the mean of the inverse of the individual antenna gains in Jy\,K$^{-1}$ provided by the NOEMA team, multiplied by the number of antennas in the phased array. 

The effective aperture efficiency of 55\,\% for individual NOEMA antennas reported in Table~\ref{tab:multiyr_sensitivities}, includes effects such as the blockage of the subreflector and losses in the interferometric signal path. If we solve Eq.~\ref{eq:dpfusd} for the apparent aperture efficiencies using the mean DPFUs reported in the same table and the total collecting area of the array, we obtain efficiencies of 51\,\% in 2021 and 53\,\% in 2022, which agree well with the nominal efficiency value.

\subsection{\pv{}} \label{sec:PV}

The official gain curve used at \pv{} is reported by \citet{penalver12}. This gain curve is based on measurements on bright sources, observed only at night time under good weather conditions in 2011. During an EHT run the weather conditions vary and the observations are made at both day and night times. Taking this into account, for the 2017--2018 EHT ANTAB tables, we used an alternative gain curve based on efficiency measurements provided by the \pv{} team in 2017 \citep{janssen19, koay23a}. While the 2021--2022 measurements are not sufficient to derive a new gain curve, we can leverage the \pv{} historic efficiency measurements during the EHT campaigns from 2017 to 2022, to update the 2017 gain curve and obtain a more representative gain curve for typical EHT observations.

\begin{figure}[h!]
    \centering
    \includegraphics[width=\textwidth]{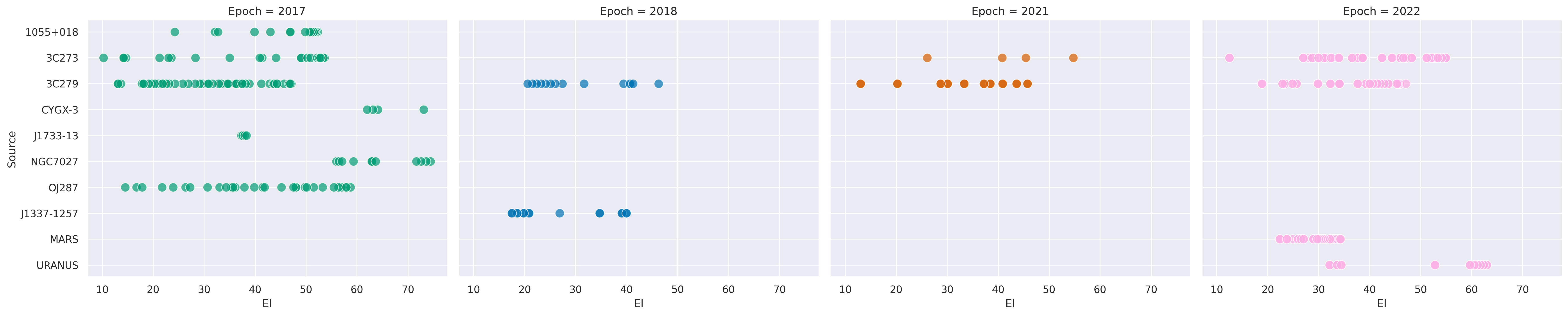}
    \caption{Elevation coverage of selected sources per year to obtain an updated \pv{} gain curve valid for typical EHT observations.}\label{fig:pv_data}
\end{figure}

\begin{figure}[h!]
    \centering
    \includegraphics[width=0.9\textwidth]{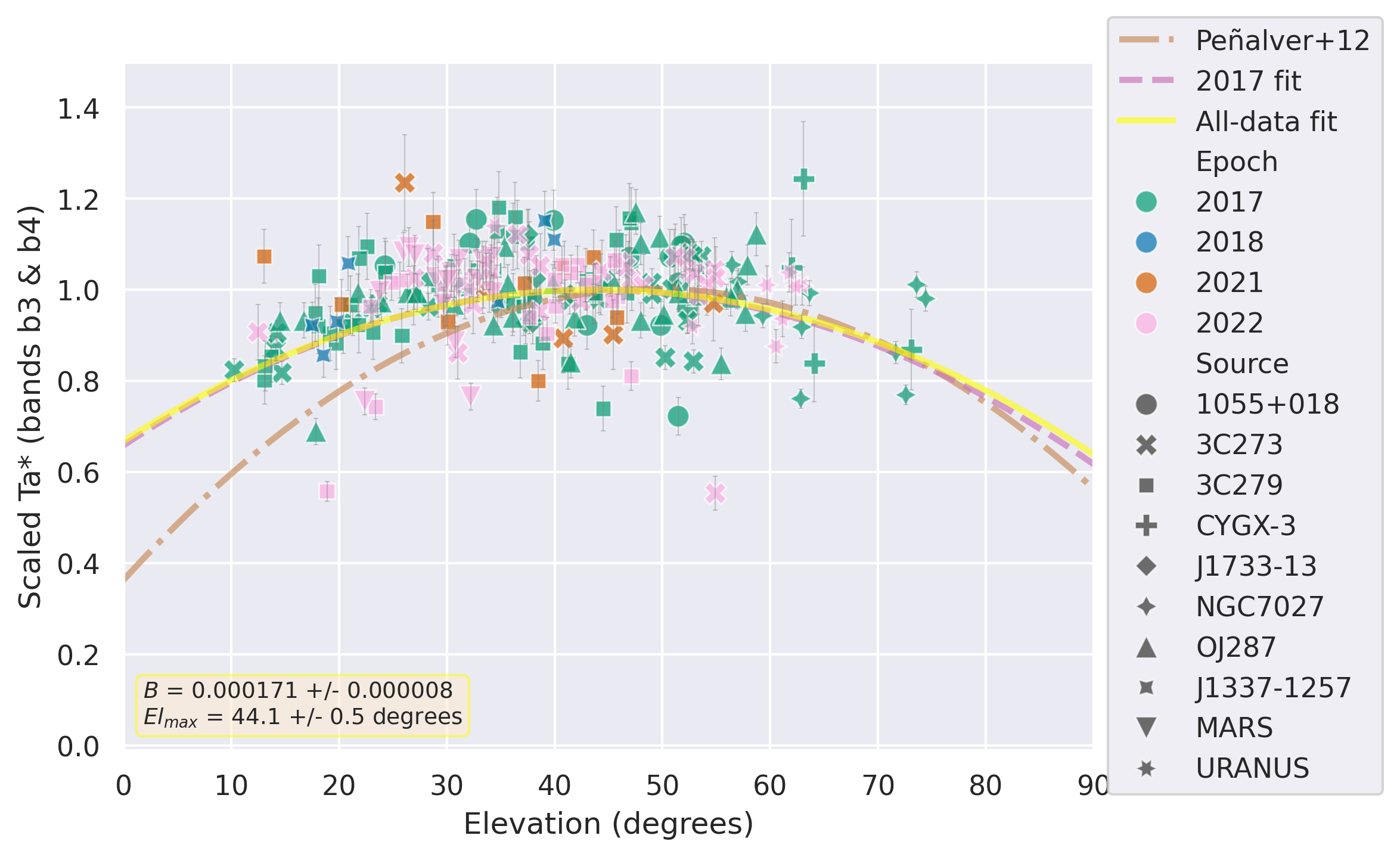}
    \caption{\pv{} antenna temperatures in USB bands measured in 2017--2022, scaled for each source in their corresponding observing epochs by their mean values within the elevation range of 40\degr{}--60\degr{}. The official \pv{} gain curve from \citet{penalver12} is represented by a dotted-dashed, dark-orange line; the dashed, dark-pink line represents the 2017 gain curve; and the solid yellow line represents the new gain curve obtained by combining data from 2017 to 2022. The box in the bottom left contains the latest gain-curve fitted parameters.}\label{fig:pv_gc}
\end{figure}

To obtain a new gain curve, we consider data that meet the following criteria:
\vspace{-0.35cm}
\begin{itemize}
    \item Scans observed with $\tau < 0.3$
    \item Sources with at least five scans, per band, per polarization, per epoch
    \item Sources with a maximum relative percent uncertainty of 15\,\%  in antenna temperature ($T^*_{\rm a}$), which corresponds to the expected uncertainty when the gain curve reaches its maximum
    \item Scans with a maximum of 15\,\% deviation from expected errors in the pointing, which are typically up to 0.1 of the half-power beam-width (HPBW) 
    \item Sources with measured full-width at half-maximum (FWHM) not more than 15\,\% above or below the HPBW, to avoid including very extended sources and/or observations with large focus errors
\end{itemize}

Such criteria resulted in the selection shown in Fig.~\ref{fig:pv_data}. Most of the historic data comes from the 2017 observations in USB bands, which covers a broad elevation range. The data in LSB bands from 2018 to 2022 is not sufficient to characterize the gain curve. Thus, we only fit the USB measurements and apply the resulting gain curve to both LSB and USB. Before the fit, we average scans made along Azimuth and Elevation for both polarizations, and then re-scale all the data using the mean of the values per source, per epoch, and per band within an elevation range of 40\degr{} to 60\degr{}. We show the final gain curve valid for 2021--2022 in Fig.~\ref{fig:pv_gc}, which is highly consistent with the one obtained for the EHT 2017 observations. The gain curve is written in its polynomial form as $g_E = 0.6677347 + 1.50811\,{\rm e}{-2}\times~E - 1.711\,{\rm e}{-4}\times~E^2$.

As previously, the \pv{} team provided 2021--2022 DPFU values per sideband, per polarization. However, these values include a correction of the official elevation-dependent gain curve from \citet{penalver12}. Since we are using a different gain curve, we obtain new DPFU estimates using the mean antenna temperatures of the selected sources (see Fig.~\ref{fig:pv_data}), divided by the corresponding flux densities provided by the \pv{} team. The new DPFU values are consistent within the uncertainties with those reported by the \pv{} team. We include these new estimates in the header of the ANTAB tables.

\subsection{SMA}\label{sec:SW}

In the 2021--2022 observing campaigns, SMA participated with an average number of six antennas\footnote{In 2021 there were six antennas participating in every track, but in 2022, the number of phased antennas varied between five and seven throughout the campaign.} (each of 6\,m in diameter), resulting in a dish with an equivalent size of $\sim15$\,m. The median phasing efficiency for primary target scans in the 2021 campaign is 99\%, and 93\% in 2022. \citet{young16} describes the beamforming system at the SMA. 

In 2021, the SMA team provided gain tables containing DPFU and sideband ratios ($r_{\rm sb}$; calculated as the ratio of the USB gain to the LSB gain) per track, per antenna, and per polarization for SMA bands 1 and 2. As in 2018, the SMA bands 1 and 2 correspond to EHT bands b3 and b4. For LSB bands, we assume that b2 is the image sideband for b3, and b1 the image sideband for b4. In 2022, the team provided modified gain tables containing values per EHT band, so no conversion between band definitions is needed.

Because of bad weather, gain information is missing on 16 April 2021 and 26--27 March 2022. By recommendation of the SMA team, for these days we use instead the gain tables provided for 17 April 2021 and 25 March 2022, respectively. Measurements on the following sources were used to estimate the DPFUs: Titan (Apr/09/2021), Vesta (Apr/13\&18/2021 and Mar/19/2022), Callisto (Apr/14--17/2021), Uranus (Mar/18\&20/2022), and Ceres (Mar/22--25/2022). These DPFUs are used to obtain the average values we report in Tables~\ref{tab:gains} and \ref{tab:multiyr_sensitivities}, after including a factor of the number of antennas in the phased array. We assume a nominal uncertainty of 5--15\,\% as in previous EHT campaigns, which is based on the dominant phase-efficiency uncertainty. We note that the sensitivities reported in Table~\ref{tab:gains} appear to be better for one polarization and worse for the other in one year, and invert the following year. While this could be related to the SMA having different receiver cartridges for RCP and LCP and the change of these over time, the difference in sensitivity arises from known system temperature calibration issues at the SMA.

We report the nominal aperture efficiency of SMA antennas in Table~\ref{tab:multiyr_sensitivities}. By solving for the apparent aperture efficiency as done for NOEMA, we obtain $\eta_{\rm A}=52.4$\,\% for the 2021--2022 EHT campaigns. This value is not used in any calculation, but it provides information of the average performance of individual SMA antennas during the EHT observations.

The SMA team provides tables per track with DSB $T^*_{\rm sys}$ measurements per minute, per antenna, per polarization. We first convert these temperatures to single-sideband (SSB) equivalent ones by using eq.~57 from \citet{issaoun17a},
\[ T^*_{\rm sys,SSB} = \left(1+r_{\rm sb}\right)\,T^*_{\rm sys,DSB},\] 
with the $r_{\rm sb}$ values provided in the gain tables. We next obtain SEFD values per antenna using Eq.~\ref{eq:sefdsd}, which are then used in  Eq.~\ref{eq:sefdph} to obtain phased SEFDs. We do this per scan, per band, and per polarization, using the nearest phasing-efficiency measurement available in the provided tables per track, containing phasing efficiency values at a cadence of 15\,s, per EHT band and per polarization.

The $T^*_{\rm sys}$ section for SMA in the ANTAB tables contains the phased SEFDs we obtained as explained above, and we thus use a DPFU$=1$ in the header. Since the gains of the SMA antennas are not elevation-dependent \citep[][]{matsushita06,koay23a}, a flat gain curve is included in the header of the ANTAB tables.

\subsection{SPT} \label{sec:SZ}

The SPT team provided tables with efficiency measurements per polarization from 221\,GHz observations to Jupiter on 14--16 April 2021, and Jupiter and Venus on 19--26 March 2022. The measurements were made at elevations $\sim$5\degr{}--15\degr{}, which are representative of the elevations at which the bulk of science targets for the EHT are observed at the SPT. 

We use an antenna diameter of 10\,m to calculate DPFU values following Eq.~\ref{eq:dpfusd}. We assume that the same DPFUs are valid across the EHT bandwidth and report the resulting values in Tables~\ref{tab:gains} and \ref{tab:multiyr_sensitivities}. As in the calibration of previous EHT observations, here we also assume a flat gain curve for the SPT given that the science targets are observed at constant elevation through a track \citep[][]{koay23a}.

\section{Metadata validation and processing pipeline update}\label{sec:pip_update}

We previously built a {\fontfamily{qcr}\selectfont python} pipeline to process and validate the EHT 2018 metadata \citep[][]{koay23a}. Since then, we have been improving the pipeline to process the metadata from new stations in the EHT array and further automate some tasks. Currently, the pipeline consists of:
\vspace{-0.35cm}
\begin{itemize}
    \item A script with functions to read the phased-array metadata from ALMA, NOEMA and SMA \citep[section~5.3 in][and Sections~\ref{sec:AA}, \ref{sec:NN} and \ref{sec:SW} in this memo]{koay23a}.
    \item A script that estimates DPFUs, elevation- and time-dependent gain curves, as well as their uncertainties (see details in Sections~\ref{sec:gains} and \ref{sec:station_meta}). The script reads gain files from the different stations when possible. We directly input the gain information from files provided by telescope representatives in non-standard format. This script automatically creates files containing ANTAB headers per band valid for all the tracks in an observing campaign, and individual files when special headers per track are needed, for example, for 14 April 2021 (see Section~\ref{sec:MG}). This section of the work was done mostly manually in the past.
    \item A main script that reads and validates the metadata from single-dish stations, and calls the two other scripts to produce validation plots, ancillary files, and ANTAB tables which include the $T^*_{\rm sys}$ or SEFDs from single-dish and phased-array stations. All these products and their corresponding README files are then automatically copied to a final metadata package for release.
\end{itemize}

For the EHT 2018 metadata release, we substituted all the missing $T^*_{\rm sys}$ values with modeled ones \citep[section 5.2.4,][]{koay23a}. However, we acknowledge that each station estimates their $T^*_{\rm sys}$ values in different ways, which may differ significantly. Hence, we now simply interpolate or extrapolate neighboring values to find missing ones. We only resort to modeling when there are no values from which to interpolate or extrapolate (e.g., GLT values from track e21e13, Section \ref{sec:GL}). 

\section{L1 Metadata Packages: Contents and Data Properties}\label{sec:l1products}

The metadata packages in this L1 release are named {\fontfamily{qcr}\selectfont EHTmetadata\_2021April\_v1.2} and {\fontfamily{qcr}\selectfont EHTmetadata\_2022March\_v1.0}, following the naming convention in \citet{koay23a}. The ANTAB tables contained in these packages use ALMA products from the first Polconversion run, using the software-based correlator {\fontfamily{qcr}\selectfont DiFX}\footnote{Distributed FX, where F stands for Fourier transform, and X for cross multiplication.} 2.7.1 and {\fontfamily{qcr}\selectfont PolConvert} 1.8.5 for 2021, and {\fontfamily{qcr}\selectfont DiFX} 2.8.1 and {\fontfamily{qcr}\selectfont PolConvert} 2.0.3 for 2022.

The 2021 metadata package is available in the public domain \url{https://almascience.eso.org/almadata/ec/eht-2021/2019.1.01218.V/group.uid___A001_X1527_X33f.ec_savolainen.metadata-ehtmetadata_2021_v1.2.tgz}. The link for the 2022 metadata is not yet available, but its public release will follow soon in the ALMA archive together with the respective data release. In Appendix~\ref{ap:ids} we provide tables that help identify participating stations and details of the observing tracks in each campaign.

We note that the ALMA archive contains three different versions of the 2021 metadata package. Version {\fontfamily{qcr}\selectfont v1.0} was released on June 2023 for early inspection of the science data. However, the ANTAB tables of this version contained data that were incorrectly labeled for two of the stations (GLT and NOEMA) and some raw data files were inadvertently missing. Because of this, a new version ({\fontfamily{qcr}\selectfont v1.1}) was released on September 2023, which is the one that was used to perform the a-priori calibration of the data presented by \citet{EHTC2025b}. After this, a new version ({\fontfamily{qcr}\selectfont v1.2}) was needed to address some issues. The main differences between versions {\fontfamily{qcr}\selectfont v1.1} and {\fontfamily{qcr}\selectfont v1.2} are:

\begin{itemize}
    \item The GLT modeled $T^*_{\rm sys}$ values for track e21e13 were obtained under incorrect assumptions in version {\fontfamily{qcr}\selectfont v1.1}. We are now using typical GLT values for $T^*_{\rm rx}$ as well as real weather conditions on 13 April from the MERRA-2 database (see Section \ref{sec:GL}). The $T^*_{\rm sys}$ values in version {\fontfamily{qcr}\selectfont v1.1} are $\sim60$\,\% larger than in {\fontfamily{qcr}\selectfont v1.2}. This means that even larger amplitude errors will remain when using version {\fontfamily{qcr}\selectfont v1.2} and these need to be considered in any downstream analysis as done in \citet{EHTC2025b}.
    \item The $T^*_{\rm sys}$ values from APEX scans where the wobbler system was enabled, do not include a correction factor in {\fontfamily{qcr}\selectfont v1.1} (see Section \ref{sec:AX}).
    \item The ANTAB tables for track e21b09 in the {\fontfamily{qcr}\selectfont v1.1} package do not indicate that APEX observed in linear polarization. 
    \item Most of the information in the {\fontfamily{qcr}\selectfont v1.1} ANTAB headers was inherited from the EHT 2018 metadata release. When comparing the mean DPFU values, the difference between versions is $<6$\,\% for APEX, SMT and JCMT, and $<1$\,\% for \pv{}. 
\end{itemize}

We expect that the gain difference between versions was absorbed by self-calibration applied to the M87$^*$ data presented in \citet{EHTC2025b}.

\subsection{Package Contents}\label{sec:contents}

The contents of the metadata package have been described in full detail in \citet{koay23a}, section~6. Here, we provide a brief description of the metadata products organized in dedicated directories. The tarball packages for 2021 and 2022, contain the following products:

\begin{itemize}
    \item \textbf{Raw metadata files} provided by telescope representatives. In some cases, we made minimal modifications to the files when formatting issues prevented the pipeline from reading them, for example, when extra information was provided in uncommented lines. These files are now in the directory {\fontfamily{qcr}\selectfont raw\_md/}.
    \item \textbf{ANTAB tables} per track and per EHT frequency band are found in the directory {\fontfamily{qcr}\selectfont antab/}. We provide two versions in their respective subdirectories as in 2018: (1) {\fontfamily{qcr}\selectfont antab/raw/} with raw tables containing the metadata provided by telescope representatives; (2) {\fontfamily{qcr}\selectfont antab/processed/} with processed tables where $T^*_{\rm sys}$ timestamps have been shifted to the middle of their respective scans,  $T^*_{\rm sys}$ outliers have been flagged, and flagged and missing $T^*_{\rm sys}$ values have been replaced via interpolation or extrapolation, or modeling when needed.
    \item \textbf{Validation plots} in the {\fontfamily{qcr}\selectfont plots/} directory. It contains plots from single-dish and phased-array stations used for quality assessment. The validation plots of the phased-arrays raw metadata are in dedicated subdirectories. We now include $T^*_{\rm sys}$ vs. time and vs. elevation plots for the single dishes after flagging and replacing missing values.
    \item \textbf{Supplementary files} produced by the pipeline, now located in the directory  {\fontfamily{qcr}\selectfont ancillary/}. It contains files that provide information of relevant issues during validation, summaries of the characteristics of the phased arrays and single-dish stations, median gains and opacities during the campaign, list of observed science sources per track, and other files with dictionaries to be used as input in downstream calibration pipelines.
\end{itemize}

Each directory contains a README file that provides a short description of each file. In the tarball package, we now include the script {\fontfamily{qcr}\selectfont custom\_antab.py} that produces customized ANTAB tables per track and per source. A README file describing the content of each directory has also been added, and when needed, it provides additional notes and a list of potential issues that may affect the a-priori amplitude of the EHT data.

\subsection{Remaining Issues in this L1 metadata release} \label{sec:dataissues}

At the time of this L1 release, there are still some peculiarities and issues in the metadata packages that need further consideration when performing the a-priori calibration of the EHT science data. 

A common issue in the 2021 and 2022 metadata packages is that the ANTAB tables do not contain ALMA $T^*_{\rm sys}$ values for scans on weak targets observed with passive phasing. These scans are not marked with the flag code ``T'' and are not easily identified in the ALMA raw metadata. The end user may need to use a nearby calibrator to obtain ALMA $T^*_{\rm sys}$ values for such scans in each track.

Here, we provide a list of the most outstanding issues per epoch that an end user should be aware of.

{\fontfamily{qcr}\selectfont EHTmetadata\_2021April\_v1.2}:
\vspace{-0.35cm}
\begin{itemize}
    \item The GLT did not provide $T^*_{\rm sys}$ values for track e21e13 owing to power-meter issues, and we included modeled values in the corresponding ANTAB tables. Analysis of the calibrated data revealed large amplitude gain errors that were mitigated by means of self-calibration \citep{EHTC2025b}. Thus, the $T^*_{\rm sys}$ values currently in the ANTAB tables are underestimated. 
    \item \pv{} reported poor pointing during track e21d15. The errors are not quantified, but we expect this to have a negative effect in the a-priori calibration.
    \item We obtained new DPFU values for \pv{} to avoid using any value inherited from their official gain curve \citep{penalver12}. However, by the time the current release was ready, it was found that the source fluxes we used to obtain the new DPFUs are obtained via their calibration on Mars, for which the official gain curve is used. The team obtained new values using the gain curve we report here, and found that the differences between DPFUs are within the uncertainties.  
    \item APEX observed in linear polarization during track e21b09. LCP corresponds to X and RCP to Y. Correct labels appear only in the ANTAB tables.

    \item During post-processing analysis, it was noted that the RCP and LCP $T^*_{\rm sys}$ values from APEX in track e21e13, band b4, might be swapped. While this issue was not confirmed by the telescope representatives, the user may need to change the labels directly in the affected ANTAB tables.
\end{itemize}
\vspace{-0.35cm}
{\fontfamily{qcr}\selectfont EHTmetadata\_2022March\_v1.0}:
\vspace{-0.35cm}
\begin{itemize}
    \item There are substantial changes to the LMT surface since 2018. The data provided in the raw metadata package is insufficient to constrain a gain curve. We have thus assumed a flat gain curve, though lacking an estimate of its uncertainty.
    \item The DPFUs for \pv{} in the ANTAB tables implicitly include the official gain curve instead of the one we report here (see related item for the 2021 package). This will be addressed in future releases.
\end{itemize}
\section{Field rotation angle dependence of EHT telescopes}\label{sec:pol}

The signal arriving at the main dish (or primary reflector) of each telescope follows a different path to the receiver, depending on the telescope's mount type, which is a combination of the drive type and the focus position.
Keeping track of this information is important if we want to disentangle the true polarization of the sources the EHT observes as an array from the transformations introduced by the optical elements in each telescope. Different orientations of the receiver antenna with respect to the celestial sky are captured by the so-called {\em field rotation angle}, and for circularly polarized feeds, the field rotation manifests as opposite phase gains of the two circularly polarized components.

The drive type for all EHT telescopes is Alt--Az, which introduces a rotation of the cabin with respect to the celestial sky as a source is being tracked. This rotation is represented by the parallactic angle $\psi_\mathrm{par}$, defined as the angle of the local zenith with respect to celestial North in the counterclockwise direction.

Regarding the focus position for an Alt--Az drive, within the EHT array there are telescopes with Cassegrain, Gregorian and Nasmyth foci configurations. In the case of a Nasmyth focus, there is an additional rotation of the receiver with respect to the optical elements fixed to the Alt--Az mount orientation that occurs as the telescope tracks in elevation. The sign of the dependence is related to the Nasmyth configuration (left or right) as well as the number of reflections prior to the Nasmyth cabin.

In addition to the parallactic angle and the elevation dependence $f_{\rm{el}}$ introduced by the focus type, we also track a possible fixed offset $\phi_\mathrm{off}$ of the receiver feed orientation with respect to its reference alignment, also measured counterclockwise on the sky (East from North).

In Table \ref{tab:frp} we provide field rotation parameters for the stations participating in the 2021 and 2022 EHT campaigns. The parameters describe a corresponding instrumental forward gain for circular feeds,
\begin{equation}
\begin{pmatrix}
g_r \\ g_l
\end{pmatrix}=
\begin{pmatrix}
e^{-i \phi} \\ e^{i \phi}
\end{pmatrix},
\quad
\text{due to field rotation angle}
\quad
\phi = f_\mathrm{par}\,\psi_\mathrm{par} + f_\mathrm{el}\, \theta_\mathrm{el} + \phi_\mathrm{off}
\end{equation}
and following the convention of \citet{EHTC2021}.

\begin{table}[ht]
\caption{Field rotation parameters for the 230\,GHz receivers at the EHT stations participating in the 2021--2022 campaigns.}\label{tab:frp}
\vspace{-0.25cm}
\begin{tabular*}{\linewidth}{@{\extracolsep{\fill}}l l c c c}
\toprule\toprule
Station & Mount type (drive, focus) & $f_{\rm{par}}$ & $f_{\rm{el}}$ & $\phi_{\mathrm{off}}$ \\
\hline 
ALMA\textsuperscript{a}   & Alt--Az, Cassegrain & 1 & 0 & $-45\degr{}$ \\    
APEX   & Alt--Az, Nasmyth-right\textsuperscript{b} & 1 & 1 & 0 \\ 
GLT    & Alt--Az, Cassegrain & 1 & 0 & 0 \\ 
\kt{}  & Alt--Az, Cassegrain\textsuperscript{c} & 1 & 0 & 0 \\ 
LMT    & Alt--Az, Nasmyth-left & 1 & $-1$ & 0 \\ 
SMT    & Alt--Az, Nasmyth-right & 1 & 1 & 0 \\  
JCMT   & Alt--Az, Cassegrain & 1 & 0 & $6.5$\degr{} \\ 
NOEMA  & Alt--Az, Cassegrain & 1 & 0 & 0 \\ 
\pv{}  & Alt--Az, Nasmyth-left & 1 & $-1$ & 0 \\ 
SMA    & Alt--Az, Nasmyth-left & 1 & $-1$ & 45\degr{} \\
SPT    & Alt--Az, Offset Cassegrain\textsuperscript{d} & 1 & 0 & -5.7\degr{} \\ 
\bottomrule
\end{tabular*}
\newline
\vspace{0.05cm}
\textsuperscript{a}\footnotesize{Details on the coordinate system for the ALMA antennas and the orientation of the receivers can be found in the ALMA Front-end Optics Design Report at \url{https://safe.nrao.edu/wiki/pub/Main/ALMAFEInfo/FEND-40.02.00.00-035-B-REPcomments-1-FP%20NT-20070320.pdf}.}\newline
\textsuperscript{b}\footnotesize{The nFLASH receiver in APEX is physically located in the Nasmyth-left cabin (left side of the telescope). However there are six reflections in the optical pathway of the signal prior to entering the Nasmyth cabin \citep{zhao2025} instead of the usual three, resulting in a flipped elevation dependence on field angle versus what is commonly found. For this reason it is labeled here as Nasmyth-right configuration.}\newline
\textsuperscript{c}\footnotesize{The receiver sits at a reimaged Cassegrain focus.}\newline
\textsuperscript{d}\footnotesize{Though the telescope is normally a Gregorian design, the VLBI system adds a pre-focus secondary mirror to create a Cassegrain system.}\newline
\end{table}

\newpage

\section*{Acknowledgements} D.G.N. acknowledges funding from Comit\'e Mixto ESO-Chile, ANID through Fondecyt Postdoctorado (project No. 3220195), and N\'ucleo Milenio TITANs (project No. NCN2022002). This work has been partly funded by the Ministry of Science and Technology (MOST, project No. 110-2124-M-001-007) and the National Science and Technology Council (NSTC, project No. 111-2124-M-001-005) from Taiwan, and the European Research Council (ERC) Synergy Grant ``BlackHolistic'' (project No. 62004280).

%%%%%%%%%%%%%%%%%%%%%%%%%%%%%%%%%%%%%%%%%

\newpage

\appendix

\counterwithin*{equation}{section}
\renewcommand\theequation{\thesection\arabic{equation}}
\setcounter{equation}{0}

%\newpage

\section{Formulae and nomenclature}\label{ap:notation}

The goal of the a-priori calibration is to scale the amplitudes measured by the telescopes to physical units. The raw metadata provided by telescope representatives contain values for the parameters we list here. We also include basic formulae that describe the relationship among them. A detailed description of basic terms and concepts for EHT and (sub-)mm VLBI flux calibration can be found in previous EHT calibration memos \citep[][and references therein]{issaoun17a,janssen19,koay23a}.

\newcolumntype{L}[1]{>{\raggedright\let\newline\\\arraybackslash}p{#1}}
\newcolumntype{C}[1]{>{\centering\let\newline\\\arraybackslash}p{#1}}
\newcolumntype{R}[1]{>{\raggedleft\let\newline\\\arraybackslash}p{#1}}

\begin{tabular}{L{0.12\textwidth}L{0.45\textwidth}R{0.38\textwidth}}
$\tau_{0}$ & Atmospheric opacity at zenith &  \\
$\tau$ & Atmospheric opacity at elevation $E$ in the line of sight & \parbox{6.5cm}{\begin{equation}\label{eq:tauE} \tau = \tau_0/{\rm \sin}(E) \end{equation}} \\ [2pt]
$T_{\rm sys}$\,(K) & System noise measured by a telescope when pointing at the empty sky, and expressed as a temperature &  \\
$T^{*}_{\rm sys}$\,(K) & Effective system noise temperature corrected for atmospheric absorption (see Eq. \ref{eq:tauE}) and other losses & \parbox{6.5cm}{\begin{equation}\label{eq:teff} T^{*}_{\rm sys} \approx e^{\tau}\,T_{\rm sys} \end{equation}} \\[17pt]
$A_{\rm geom}$\,(m$^2$) & Geometrical area of the antenna measuring the signal & \\[4pt]
$\eta_{\rm A}$\,(\%) & Aperture efficiency (it accounts for the surface accuracy of the antenna) & \\[16pt]
$\eta_{\rm ph}$\,(\%) & Phasing efficiency (it characterizes the loss of effective collecting area by an array) & \\[-5pt]
DPFU$_{\rm sd}$\,(K\,Jy$^{-1}$) & Degrees per flux unit for a single dish telescope ($k_{\rm B}$ is the Boltzmann constant) & \parbox{6.5cm}{\begin{equation}\label{eq:dpfusd}  {\rm DPFU}_{\rm sd} = \eta_{\rm A}\,\frac{A_{\rm geom}}{2k_{\rm B}} \end{equation}} \\[2pt]
DPFU$_{\rm ph}$\,(K\,Jy$^{-1}$) & Degrees per flux unit for an array of $N_{\rm ant}$ antennas & \parbox{6.5cm}{\begin{equation}\label{eq:dpfuph} {\rm DPFU}_{\rm ph} = 
{\rm DPFU}_{\rm sd}\,N_{\rm ant} \end{equation}} \\[4pt]
$g_E$ & Elevation dependent gain curve (see equations \ref{eq:gc_poly} and \ref{eq:gc_alt}) & \\[-6pt]
SEFD$_{\rm sd}$\,(Jy) & System equivalent flux density for a single dish telescope & \parbox{6.5cm}{\begin{equation}\label{eq:sefdsd} {\rm SEFD}_{\rm sd} = \frac{T^{*}_{\rm sys}}{g_E\,{\rm DPFU}_{\rm sd}} \end{equation}} \\[-16pt]
SEFD$_{\rm ph}$\,(Jy) & System equivalent flux density for an array of $N_{\rm ant}$ antennas \citep[e.g.,][]{dewey94}& \parbox{6.5cm}{\begin{equation}\label{eq:sefdph} {\rm SEFD}_{\rm ph} = \left(\eta_{\rm ph} \,\sum_{i=1}^{N_{\rm ant}}\left(\frac{1}{{\rm SEFD}_{\rm i}}\right)\right)^{-1} \end{equation}}
\end{tabular}

Note that normally DPFU$_{\rm ph}$ includes the phasing efficiency factor ($\eta_{\rm ph}$), but here we include it in SEFD$_{\rm ph}$ instead.

\newpage
\section{Naming Conventions and Identifications for Stations and Tracks}\label{ap:ids}

There are different naming conventions for telescopes and for observing tracks which may be preferred by different calibration pipelines and plotting libraries. As in \citet{koay23a}, we provide a list for each in Tables~\ref{tab:stations} and \ref{tab:tracks}, respectively. Here, VEX stands for ``VLBI Experiment''\footnote{\url{https://vlbi.org/wp-content/uploads/2019/03/vex-definition-15b1.pdf}}, and HOPS stands for the ``Haystack Observatory Postprocessing System''\footnote{\url{https://www.haystack.mit.edu/haystack-observatory-postprocessing-system-hops/}} software.

\begin{table}[ht]
\caption{Names and identification codes for all the telescopes that have participated in EHT science observations in 2021 and 2022.}\label{tab:stations}
\vspace{-0.25cm}
\begin{tabular*}{\linewidth}{@{\extracolsep{\fill}}l c c c c}
\toprule\toprule
Full Telescope Name & Abbreviation & VEX Site Name & VEX Code & HOPS Code\\
\hline 
Atacama Large Millimeter/submillimeter Array &  ALMA & ALMA & Aa & A\\    
Atacama Pathfinder Experiment  & APEX & APEX & Ax & X \\
Greenland Telescope & GLT & THULE & Gl & G \\
University of Arizona 12-meter Telescope & \kt{} & KITTPEAK & Kt\textsuperscript{a} & K \\
Large Millimeter Telescope Alfonso Serrano & LMT & LMT & Lm & L\\
Submillimeter Telescope & SMT & SMTO & Mg & Z\\
James Clerk Maxwell Telescope & JCMT & JCMT & Mm & J\\
NOrthern Extended Millimeter Array & NOEMA & NOEMA & Nn & N \\
Institut de Radioastronomie Millim\'{e}trique 30\,m Telescope & IRAM 30\,m & PICOVEL & Pv & P \\
Submillimeter Array & SMA & SMAP & Sw & S\\
South Pole Telescope & SPT & SPT & Sz & Y\\
\bottomrule
\end{tabular*}
\newline
\vspace{0.05cm}
\textsuperscript{a}\footnotesize{\kt{} is also referred to as ``Kp'' \citep[e.g.][]{EHTC2025b}.}\newline
\end{table}

\begin{table}[ht]
\caption{VEX and EHT experiment identification numbers for each track of the 2021 and 2022 EHT observing campaigns. Participating stations for each track are also shown.}\label{tab:tracks}
\vspace{-0.25cm}
\begin{tabular*}{\linewidth}{@{\extracolsep{\fill}}l c c l}
\toprule\toprule
Observing Date\textsuperscript{a} (UT) & VEX ID & EHT Experiment No. & Participating Stations\textsuperscript{b}  \\
\hline 
Apr/09/2021 & e21b09 &	3762 & \textit{Aa}, Ax, Gl, Kt, Mg, Mm, Nn, \textit{Pv}, Sw, Sz \\ 
Apr/13/2021 & e21e13 &	3763 & Aa, Ax, Gl, Kt, Mg, Mm, \textit{Nn}, Pv, Sw, Sz \\ 
Apr/14/2021 & e21a14 & 	3764 & Aa, Ax, Gl, Kt, Mg, Mm, Nn, Pv, Sw, Sz \\ 
Apr/15/2021 & e21d15 &	3765 & Aa, Ax, Gl, Kt, Mg, Mm, Nn, Pv, Sw, Sz \\ 
Apr/16/2021 & e21a16 & 	3766 & \textit{Aa}, \textit{Ax}, \textit{Gl}, Kt, Mg, Mm, \textit{Nn}, \textit{Pv}, Sw, Sz \\ 
Apr/17/2021 & e21a17 & 	3767 & Aa, Ax, Gl, Kt, Mg, Mm, Nn, Pv, Sw, Sz \\ 
Apr/18/2021 & e21e18 &	3768 & Aa, Ax, Gl, Kt, Mg, Mm, Nn, Pv, Sw, \textit{Sz} \\
\midrule
Mar/18/2022 & e22g18 & 	3803 & \textit{Aa}, \textit{Ax}, Gl, Kt, Lm, Mg, Mm, Nn, Pv, Sw \\
Mar/19/2022 & e22b19 & 	3804 & Aa, Ax, Gl, Kt, Lm, Mg, Mm, Nn, Pv, Sw, Sz \\
Mar/20/2022 & e22c20 & 	3805 & Aa, Ax, Gl, Kt, Lm, Mg, Mm, Nn, Pv, Sw, Sz \\
Mar/22/2022 & e22e22 & 	3806 & Aa, Ax, Gl, Kt, Lm, Mg, Mm, Nn, Pv, Sw, Sz \\
Mar/23/2022 & e22d23 & 	3807 & Aa, Ax, Kt, Lm, Mg, Mm, Nn, \textit{Pv}, Sw, Sz \\
Mar/26/2022 & e22a26 & 	3808 & Aa, Ax, Gl, Kt, Lm, Mg, Mm, Nn, Pv, Sw, Sz \\
Mar/27/2022 & e22f27 & 	3809 & Aa, Ax, Gl, Kt, Lm, Mg, Mm, Nn, Pv, Sw \\
\bottomrule
\end{tabular*}
\newline
\vspace{0.05cm}
\textsuperscript{a}\footnotesize{The bulk of the observations for a given track were performed on the listed date. In some cases, the first scans have timestamps corresponding to the previous UT day.}\\
\textsuperscript{b}\footnotesize{Stations that could not observe owing to bad weather or technical issues are shown in italics. Note that the listed stations are the ones that were scheduled for a given track, and not all the sources in a track were necessarily observed by all of them.}\newline
\end{table}

\clearpage


\begin{thebibliography}{}
\bibitem[Butler(2012)]{butler12} Butler, B. 2012, ALMA Memo 594
\bibitem[Chen et al.(2023)]{chen23} Chen M.-T., Asada K., Matsushita S., Raffin P., Inoue M., Ho P.~T.~P., Han C.-C., et al., 2023, PASP, 135, 095001
\bibitem[Crew et al.(2023)]{crew23} Crew, G.~B., Goddi, C., Matthews, L.~D., et al.\ 2023, \pasp, 135, 025002
\bibitem[Dewey(1994)]{dewey94} Dewey, R.~J.\ 1994, \aj, The Effect of Correlated Noise in Phased-Array Observations of Radio Sources, 108, 337 %doi:10.1086/117073
\bibitem[Drain and Gough(1996)]{drain96} Drain, D. and Gough, A.~M.\ 1996, IEEE Transactions on Semiconductor Manufacturing, Applications of the upside-down normal loss function, 9, 143-145 %doi:10.1109/66.484295
\bibitem[Event Horizon Telescope Collaboration et al.(2019)]{EHTC2019c} Event Horizon Telescope Collaboration, et al., 2019, ApJL, 875, L3
\bibitem[Event Horizon Telescope Collaboration et al.(2021)]{EHTC2021} Event Horizon Telescope Collaboration, et al., 2021, ApJL, 910, L12
\bibitem[Event Horizon Telescope Collaboration et al.(2025)]{EHTC2025b} Event Horizon Telescope Collaboration, et al., 2025, accepted for publication in A\&A
\bibitem[Gelaro et al.(2017)]{gelaro17} Gelaro, R., McCarty, W., Su{\'a}rez, M.~J., et al.\ 2017, Journal of Climate, 30, 14, 5419 %doi:10.1175/JCLI-D-16-0758.1
\bibitem[Goddi et al.(2019)]{goddi19} Goddi, C., Mart\'{\i}-Vidal I., Messias, H., et al. 2019, \pasp, 131, 075003 
\bibitem[Goddi et al.(2025)]{goddi25} Goddi, C., Carlos, D.~F., Crew, G.~B., et al. 2025, arXiv:2505.10181 %doi:10.48550/arXiv.2505.10181
\bibitem[Issaoun et al.(2017a)]{issaoun17a} Issaoun S., Folkers, T. W., Blackburn, et al. 2017, EHT memo series, 2017-CE-02 \\ \url{https://eventhorizontelescope.org/sites/g/files/omnuum3116/files/eht/files/EHT_memo_Issaoun_2017-CE-02.pdf}
\bibitem[Issaoun et al.(2017b)]{issaoun17b} Issaoun S., Folkers, T. W., Marrone, D. P., Kim, J., Tilanus, R., and Falcke, H. 2017, EHT memo series, 2017-CE-03 \url{https://eventhorizontelescope.org/sites/g/files/omnuum3116/files/eht/files/EHT_memo_Issaoun_2017-CE-03.pdf}
\bibitem[Issaoun et al.(2018)]{issaoun18} Issaoun S., Falcke, H., Friberg, P., Janssen, M., Tilanus, R., and Wouterloot, J. 2018, EHT memo series, 2018-CE-01 \url{https://eventhorizontelescope.org/sites/g/files/omnuum3116/files/eht/files/EHT_memo_Issaoun_2018-CE-01.pdf}
\bibitem[Janssen et al.(2019)]{janssen19} Janssen, M., Blackburn, L., Issaoun, S., Krichbaum, T. P., Wielgus, M. 2019, EHT memo series, 2019-CE-01 \\ \url{https://eventhorizontelescope.org/sites/g/files/omnuum3116/files/eht/files/EHT_memo_Janssen_2019-CE-01.pdf}
\bibitem[Keenan(2023)]{keenan23} Keenan, R.~P.\ 2023, Ph.D. Thesis, Tools for Measuring the Cosmic History of Molecular Gas, University of Arizona. 
\bibitem[Koay et al.(2023a)]{koay23a} Koay, J.~Y., Romero-Ca{\~n}izales, C., Matthews, L.~D., et al.\ 2023, EHT memo series, 2023-L1-01 \\ \url{https://eventhorizontelescope.org/sites/g/files/omnuum3116/files/eht/files/eht_memo_koay_2023-l1-01.pdf}
\bibitem[Koay et al.(2023b)]{koay23b} Koay, J.~Y., Asada, K., Matsushita, S., et al.\ 2023, EHT memo series, 2023-L1-02, \\ \url{https://eventhorizontelescope.org/sites/g/files/omnuum3116/files/eht/files/eht_memo_koay_2023-l1-02.pdf}
\bibitem[Mart\'{\i}-Vidal et al.(2016)]{martividal16} Mart\'{\i}-Vidal I., Roy, A., Conway, J., \& Zensus, A. J. 2016, A\&A, 587, A143
\bibitem[Matsushita et al.(2006)]{matsushita06}Matsushita, S., Saito, M., Sakamoto, K., et al. 2006, in Proceedings of the SPIE, Vol. 6275, Society of Photo-Optical Instru- mentation Engineers (SPIE) Conference Series, 62751W
\bibitem[Matthews et al.(2018)]{matthews18} Matthews, L.~D., Crew, G.~B., Doeleman, S.~S., et al.\ 2018, \pasp, 130, 015002 %doi:10.1088/1538-3873/aa9c3d
\bibitem[McMullin et al.(2007)]{mcmullin07} McMullin, J. P., Waters, B., Schiebel, D., Young, W., and Golap, K. 2007, in Shaw, R. A., Hill, F., Bell, D. J. eds, ASP Conference Series, Vol. 376, Astronomical Data Analysis Software and Systems XVI. p. 127
\bibitem[Mizuno et al.(2020)]{mizuno20} Mizuno, I., Friberg, P., Berthold, R., et al.\ 2020, \procspie, 11453, 114533T. %doi:10.1117/12.2561742
\bibitem[Mizuno \& Han(2021)]{mizuno21} Mizuno, I. \& Han, C.-C.\ 2021, Nature Astronomy, 5, 331. %doi:10.1038/s41550-021-01324-0
\bibitem[Molod et al.(2015)]{molod15} Molod, A., Takacs, L., Suarez, M., et al.\ 2015, Geoscientific Model Development, 8, 5, 1339 %doi:10.5194/gmd-8-1339-201510.5194/gmdd-7-7575-2014
\bibitem[Paine, S.(2022)]{paine22} Paine, S. 2022, The am atmospheric model (12.2). Zenodo %\url{https://doi.org/10.5281/zenodo.6774378}
\bibitem[Pe\~nalver(2012)]{penalver12} Pe\~nalver, J. 2012, IRAM Technical Report, Antenna Technical Works. Institut de Radio Astronomie Millimetrique, Saint-Martin-d'H\`eres, France
\bibitem[Pesce et al.(2024)]{pesce24} Pesce, D.~W., Blackburn, L., Chaves, R., et al.\ 2024, ApJ, 968, 2, 69 %doi:10.3847/1538-4357/ad3961
\bibitem[Pi{\'e}tu et al.(2025)]{pietu25} Pi{\'e}tu, V., Garc{\'\i}a, R., Brogui{\`e}re, D., et al.\ 2025, A\&A, 704, A86. %doi:10.1051/0004-6361/202556497
\bibitem[Rienecker et al.(2011)]{rienecker11} Rienecker, M.~M., Suarez, M.~J., Gelaro, R., et al.\ 2011, Journal of Climate, 24, 14, 3624 %doi:10.1175/JCLI-D-11-00015.1
\bibitem[von Fellenberg et al.(2025)]{vonfellenberg25} von Fellenberg, S.~D., Garc{\'\i}a, R., Janssen, M., et al.\ 2025, Research Notes of the American Astronomical Society, 9, 6, 134 %doi:10.3847/2515-5172/addf47
\bibitem[Young et al.(2016)]{young16} Young, A., Primiani, R., Weintroub, J., et al.\ 2016, Phased Array Systems and Technology (ARRAY, Performance assessment of an adaptive beamformer for the submillimeter array, E1 %doi:10.1109/ARRAY.2016.7832576
\bibitem[Zhao et al.(2025)]{zhao2025} Zhao G.-Y., Roy A.~L., Wagner J.~F., Donoso E., Torne P., Ros E., Lindqvist M., et al., 2025, arXiv, arXiv:2507.10190
\end{thebibliography}
\end{document}